\pdfoutput=1
\documentclass[12pt,a4paper]{article}

\usepackage{subcaption}
\usepackage{ifthen} 
\newboolean{pdflatex}
\setboolean{pdflatex}{true} 

\newboolean{articletitles}
\setboolean{articletitles}{true} 

\newboolean{uprightparticles}
\setboolean{uprightparticles}{false} 

\def\paperauthors{LHCb collaboration} 
\def\paperasciititle{Measurement of associated J/psi-Psi2S production cross-section in pp collisions at sqrt(s)=13 TeV} 
\def\papertitle{Measurement of associated \jpsi-\psitwos production cross-section in $pp$ collisions at $\sqs=13\tev$} 
\def\paperkeywords{{High Energy Physics}, {LHCb}} 
\def\papercopyright{\the\year\ CERN for the benefit of the LHCb collaboration} 
\def\paperlicence{CC BY 4.0 licence}
\def\paperlicenceurl{https://creativecommons.org/licenses/by/4.0/}

\usepackage[top=1in, bottom=1.25in, left=1in, right=1in]{geometry}

\usepackage{microtype}
\usepackage{lineno}  
\usepackage{xspace} 
\usepackage{caption} 

\usepackage{graphicx}  
\usepackage{color}
\usepackage{colortbl}
\graphicspath{{./figs/}} 

\usepackage{amsmath} 
\usepackage{amssymb}
\usepackage{amsfonts}
\usepackage{upgreek} 

\newcommand*\patchAmsMathEnvironmentForLineno[1]{%
\expandafter\let\csname old#1\expandafter\endcsname\csname #1\endcsname
\expandafter\let\csname oldend#1\expandafter\endcsname\csname
end#1\endcsname
 \renewenvironment{#1}%
   {\linenomath\csname old#1\endcsname}%
   {\csname oldend#1\endcsname\endlinenomath}%
}
\newcommand*\patchBothAmsMathEnvironmentsForLineno[1]{%
  \patchAmsMathEnvironmentForLineno{#1}%
  \patchAmsMathEnvironmentForLineno{#1*}%
}
\AtBeginDocument{%
\patchBothAmsMathEnvironmentsForLineno{equation}%
\patchBothAmsMathEnvironmentsForLineno{align}%
\patchBothAmsMathEnvironmentsForLineno{flalign}%
\patchBothAmsMathEnvironmentsForLineno{alignat}%
\patchBothAmsMathEnvironmentsForLineno{gather}%
\patchBothAmsMathEnvironmentsForLineno{multline}%
\patchBothAmsMathEnvironmentsForLineno{eqnarray}%
}

\usepackage{hyperxmp}

\usepackage[pdftex,
            pdfauthor={\paperauthors},
            pdftitle={\paperasciititle},
            pdfkeywords={\paperkeywords},
            pdfcopyright={Copyright (C) \papercopyright},
            pdflicenseurl={\paperlicenceurl}]{hyperref}

\usepackage[colorinlistoftodos,textsize=scriptsize]{todonotes}

\usepackage[bottom,flushmargin,hang,multiple]{footmisc}

\usepackage[all]{hypcap} 

\usepackage{xspace} 
\usepackage{upgreek}

\def\lhcb   {\mbox{LHCb}\xspace}

\def\MagUp {\mbox{\em Mag\kern -0.05em Up}\xspace}

\ifthenelse{\boolean{uprightparticles}}%
{

 \def\Pmu         {\ensuremath{\upmu}\xspace}

 \def\Pchi        {\ensuremath{\upchi}\xspace}                 
 \def\Ppsi        {\ensuremath{\uppsi}\xspace}

 \def\PDelta      {\ensuremath{\Delta}\xspace}                 
 \def\PXi         {\ensuremath{\Xi}\xspace}                 
 \def\PLambda     {\ensuremath{\Lambda}\xspace}                 
 \def\PSigma      {\ensuremath{\Sigma}\xspace}                 
 \def\POmega      {\ensuremath{\Omega}\xspace}                 
 \def\PUpsilon    {\ensuremath{\Upsilon}\xspace}
 \let\oldPi\Pi
 \def\PPi         {\ensuremath{\oldPi}\xspace}

 \def\PB      {\ensuremath{\mathrm{B}}\xspace}                 
                  
 \def\PD      {\ensuremath{\mathrm{D}}\xspace}

 \def\PJ      {\ensuremath{\mathrm{J}}\xspace}                 
 \def\PK      {\ensuremath{\mathrm{K}}\xspace}

 \def\Pb      {\ensuremath{\mathrm{b}}\xspace}                 
 \def\Pc      {\ensuremath{\mathrm{c}}\xspace}                 
                  
 \def\Pe      {\ensuremath{\mathrm{e}}\xspace}

 \def\Pi      {\ensuremath{\mathrm{i}}\xspace}

 \def\Pp      {\ensuremath{\mathrm{p}}\xspace}

 \def\Ps      {\ensuremath{\mathrm{s}}\xspace}

 \def\thebaroffset{0.0em}
}
{

 \def\Pmu         {\ensuremath{\mu}\xspace}

 \def\Pchi        {\ensuremath{\chi}\xspace}                 
 \def\Ppsi        {\ensuremath{\psi}\xspace}                 
                  
 \mathchardef\PDelta="7101
 \mathchardef\PXi="7104
 \mathchardef\PLambda="7103
 \mathchardef\PSigma="7106
 \mathchardef\POmega="710A
 \mathchardef\PUpsilon="7107
 \mathchardef\PPi="7105
                  
 \def\PB      {\ensuremath{B}\xspace}                 
                  
 \def\PD      {\ensuremath{D}\xspace}

 \def\PJ      {\ensuremath{J}\xspace}                 
 \def\PK      {\ensuremath{K}\xspace}

 \def\Pb      {\ensuremath{b}\xspace}                 
 \def\Pc      {\ensuremath{c}\xspace}                 
                  
 \def\Pe      {\ensuremath{e}\xspace}

 \def\Pi      {\ensuremath{i}\xspace}

 \def\Pp      {\ensuremath{p}\xspace}

 \def\Ps      {\ensuremath{s}\xspace}

 \def\thebaroffset{0.18em}
}
\newcommand{\offsetoverline}[2][\thebaroffset]{\kern #1\overline{\kern -#1 #2}}%

\makeatletter
\ifcase \@ptsize \relax
  \newcommand{\miniscule}{\@setfontsize\miniscule{4}{5}}
\or
  \newcommand{\miniscule}{\@setfontsize\miniscule{5}{6}}
\or
  \newcommand{\miniscule}{\@setfontsize\miniscule{5}{6}}
\fi
\makeatother

\DeclareRobustCommand{\optbar}[1]{\shortstack{{\miniscule (\rule[.5ex]{1.25em}{.18mm})}
  \\ [-.7ex] $#1$}}

\def\epem       {{\ensuremath{\Pe^+\Pe^-}}\xspace}

\def\mumu       {{\ensuremath{\Pmu^+\Pmu^-}}\xspace}

\def\squark    {{\ensuremath{\Ps}}\xspace}

\def\cquark    {{\ensuremath{\Pc}}\xspace}

\def\bquark    {{\ensuremath{\Pb}}\xspace}
\def\bquarkbar {{\ensuremath{\overline \bquark}}\xspace}
\def\bbbar     {{\ensuremath{\bquark\bquarkbar}}\xspace}

\def\KorKbar {\kern \thebaroffset\optbar{\kern -\thebaroffset \PK}{}\xspace}

\def\D       {{\ensuremath{\PD}}\xspace}

\def\DorDbar {\kern \thebaroffset\optbar{\kern -\thebaroffset \PD}\xspace}

\def\Dp      {{\ensuremath{\D^+}}\xspace}
\def\Dm      {{\ensuremath{\D^-}}\xspace}

\def\DpDm    {\ensuremath{\Dp {\kern -0.16em \Dm}}\xspace}

\def\B       {{\ensuremath{\PB}}\xspace}

\def\BorBbar {\kern \thebaroffset\optbar{\kern -\thebaroffset \PB}\xspace}

\def\Bd      {{\ensuremath{\B^0}}\xspace}

\def\BdorBdbar {\kern \thebaroffset\optbar{\kern -\thebaroffset \Bd}\xspace}

\def\Bs      {{\ensuremath{\B^0_\squark}}\xspace}

\def\BsorBsbar {\kern \thebaroffset\optbar{\kern -\thebaroffset \Bs}\xspace}

\def\jpsi     {{\ensuremath{{\PJ\mskip -3mu/\mskip -2mu\Ppsi}}}\xspace}
\def\psitwos  {{\ensuremath{\Ppsi{(2S)}}}\xspace}

\def\chic     {{\ensuremath{\Pchi_\cquark}}\xspace}

\def\Upsilonres  {{\ensuremath{\PUpsilon}}\xspace}
\def\Y#1S{\ensuremath{\PUpsilon{(#1S)}}\xspace}

\def\proton      {{\ensuremath{\Pp}}\xspace}
\def\antiproton  {{\ensuremath{\overline \proton}}\xspace}

\def\LorLbar     {\kern \thebaroffset\optbar{\kern -\thebaroffset \PLambda}\xspace}

\def\BF         {{\ensuremath{\mathcal{B}}}\xspace}
\def\BR         {\BF}

\def\to                 {\ensuremath{\rightarrow}\xspace}

\def\eps   {{\ensuremath{\varepsilon}}\xspace}

\newcommand{\etot}{\ensuremath{\varepsilon^{\mathrm{ tot}}}\xspace}
\def\eacc {\ensuremath{\eps^{\rm acc}}\xspace}
\def\erecsel {\ensuremath{\eps^{\rm rec\&sel}}\xspace}
\def\etri {\ensuremath{\eps^{\rm trig}}\xspace}
\def\epid {\ensuremath{\eps^{\rm PID}}\xspace}

\def\AT#1     {\ensuremath{A_{\mathrm{T}}^{#1}}\xspace}           

\def\C#1      {\ensuremath{\mathcal{C}_{#1}}\xspace}                       
\def\Cp#1     {\ensuremath{\mathcal{C}_{#1}^{'}}\xspace}                    
\def\Ceff#1   {\ensuremath{\mathcal{C}_{#1}^{\mathrm{(eff)}}}\xspace}        
\def\Cpeff#1  {\ensuremath{\mathcal{C}_{#1}^{'\mathrm{(eff)}}}\xspace}       
\def\Ope#1    {\ensuremath{\mathcal{O}_{#1}}\xspace}                       
\def\Opep#1   {\ensuremath{\mathcal{O}_{#1}^{'}}\xspace}                    

\newcommand{\nospaceunit}[1]{\ensuremath{\text{#1}}}       
\newcommand{\aunit}[1]{\ensuremath{\text{\,#1}}}       

\newcommand{\tev}{\aunit{Te\kern -0.1em V}\xspace}
\newcommand{\gev}{\aunit{Ge\kern -0.1em V}\xspace}
\newcommand{\mev}{\aunit{Me\kern -0.1em V}\xspace}
\newcommand{\kev}{\aunit{ke\kern -0.1em V}\xspace}
\newcommand{\ev}{\aunit{e\kern -0.1em V}\xspace}
 
\newcommand{\mevc}{\ensuremath{\aunit{Me\kern -0.1em V\!/}c}\xspace}
\newcommand{\gevc}{\ensuremath{\aunit{Ge\kern -0.1em V\!/}c}\xspace}
\newcommand{\mevcc}{\ensuremath{\aunit{Me\kern -0.1em V\!/}c^2}\xspace}
\newcommand{\gevcc}{\ensuremath{\aunit{Ge\kern -0.1em V\!/}c^2}\xspace}

\def\mum  {\ensuremath{\,\upmu\nospaceunit{m}}\xspace}

\def\mbarn{\aunit{mb}\xspace}
\def\mub{\ensuremath{\,\upmu\nospaceunit{b}}\xspace}
\def\nb {\aunit{nb}\xspace}

\def\fb   {\ensuremath{\aunit{fb}}\xspace}
\def\invfb   {\ensuremath{\fb^{-1}}\xspace}

\def\ps   {\ensuremath{\aunit{ps}}\xspace}

\newcommand{\chisqip}{\ensuremath{\chi^2_{\text{IP}}}\xspace}

\def\deriv {\ensuremath{\mathrm{d}}}

\def\gsim{{~\raise.15em\hbox{$>$}\kern-.85em
          \lower.35em\hbox{$\sim$}~}\xspace}
\def\lsim{{~\raise.15em\hbox{$<$}\kern-.85em
          \lower.35em\hbox{$\sim$}~}\xspace}

\def\sPlot{\mbox{\em sPlot}\xspace}

\def\sWeight{\mbox{\em sWeight}\xspace}
\def\sWeights{\mbox{\em sWeights}\xspace}

\def\sqs   {\ensuremath{\protect\sqrt{s}}\xspace}

\def\pt         {\ensuremath{p_{\mathrm{T}}}\xspace}

\def\ptot       {\ensuremath{p}\xspace}

\newcommand{\lum} {\ensuremath{\mathcal{L}}\xspace}

\def\evtgen     {\mbox{\textsc{EvtGen}}\xspace}

\def\geant      {\mbox{\textsc{Geant4}}\xspace}

\def\photos     {\mbox{\textsc{Photos}}\xspace}

\def\pythia     {\mbox{\textsc{Pythia}}\xspace}

\def\tell1  {TELL1\xspace}
\def\ukl1   {UKL1\xspace}

\newcommand{\eg}{\mbox{\itshape e.g.}\xspace}
\newcommand{\ie}{\mbox{\itshape i.e.}\xspace}

\newcommand{\lhcborcid}[1]{\href{https://orcid.org/#1}{\hspace*{0.1em}\raisebox{-0.45ex}{\includegraphics[width=1em]{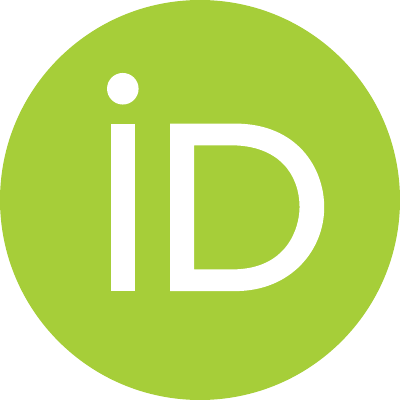}}}}

\usepackage{cite} 
\usepackage{mciteplus}

\usepackage{longtable} 
\usepackage{multirow}
\usepackage[utf8]{inputenc}

\begin{document}

\renewcommand{\thefootnote}{\fnsymbol{footnote}}
\setcounter{footnote}{1}


\begin{titlepage}
\pagenumbering{roman}

\vspace*{-1.5cm}
\centerline{\large EUROPEAN ORGANIZATION FOR NUCLEAR RESEARCH (CERN)}
\vspace*{1.5cm}
\noindent
\begin{tabular*}{\linewidth}{lc@{\extracolsep{\fill}}r@{\extracolsep{0pt}}}
\ifthenelse{\boolean{pdflatex}}
{\vspace*{-1.5cm}\mbox{\!\!\!\includegraphics[width=.14\textwidth]{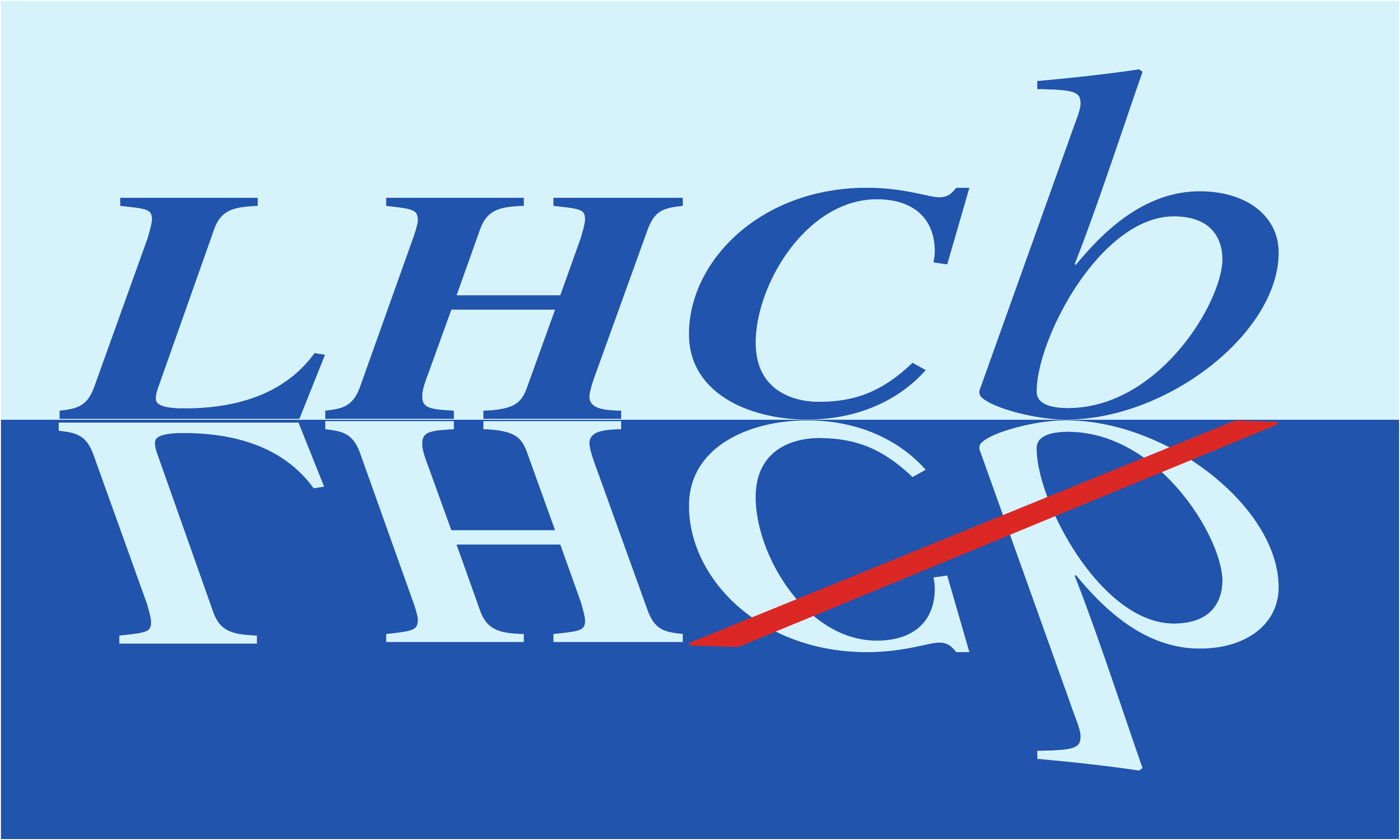}} & &}%
{\vspace*{-1.2cm}\mbox{\!\!\!\includegraphics[width=.12\textwidth]{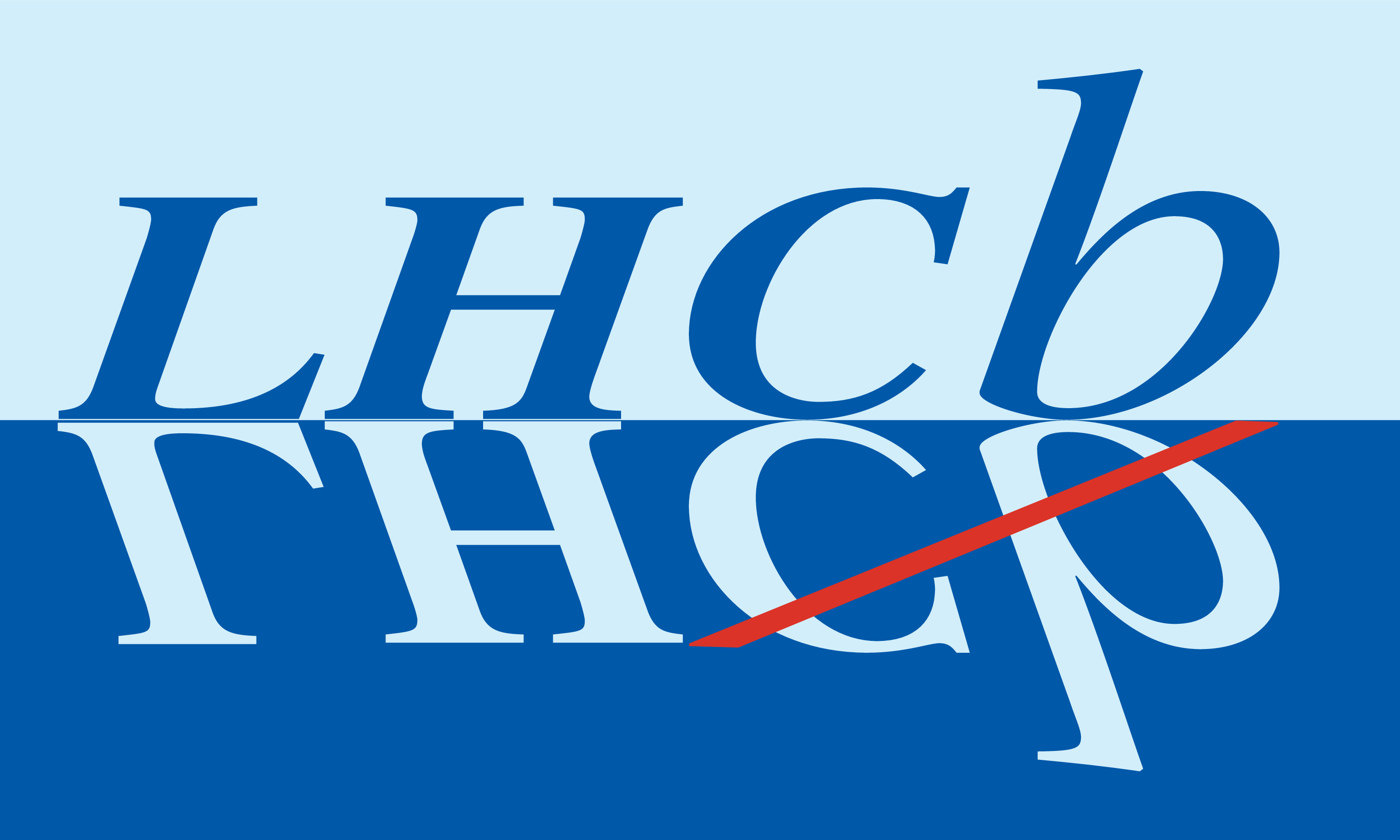}} & &}%
\\
 & & CERN-EP-2023-243 \\  
 & & LHCb-PAPER-2023-023 \\  
 & & May 23, 2024 \\ 
\end{tabular*}

\vspace*{4.0cm}

{\normalfont\bfseries\boldmath\huge
\begin{center}
  \papertitle 
\end{center}
}

\vspace*{1.0cm}

\begin{center}
\paperauthors\footnote{Authors are listed at the end of this paper.}
\end{center}

\vspace{\fill}

\begin{abstract}
  \noindent
    The cross-section of associated \jpsi-\psitwos production in proton-proton collisions at a centre-of-mass energy of $\sqs=13\tev$ is measured using a data sample corresponding to an integrated luminosity of $4.2\invfb$, collected by the \lhcb experiment.
    The measurement is performed for both \jpsi and \psitwos mesons having transverse momentum ${\pt<14\gevc}$ and rapidity ${2.0<y<4.5}$, assuming negligible polarisation of the \jpsi and \psitwos mesons.
    The production cross-section is measured to be \mbox{$4.5\pm0.7\pm0.3\nb$}, where the first uncertainty is statistical and the second systematic.
    The differential cross-sections are measured as functions of several kinematic variables of the \jpsi-\psitwos candidates.
    The results are combined with a measurement of \jpsi-\jpsi production, giving a cross-section ratio between \jpsi-\psitwos and \jpsi-\jpsi production of \mbox{$0.274\pm0.044\pm0.008$}, where the first uncertainty is statistical and the second systematic.
  
\end{abstract}

\vspace*{1.0cm}

\begin{center}
  Published in JHEP 05 (2024) 259
\end{center}

\vspace{\fill}

{\footnotesize 
\centerline{\copyright~\papercopyright. \href{\paperlicenceurl}{\paperlicence}.}}
\vspace*{2mm}

\end{titlepage}


\newpage
\setcounter{page}{2}
\mbox{~}
%
%
%
%


\renewcommand{\thefootnote}{\arabic{footnote}}
\setcounter{footnote}{0}

\cleardoublepage


\pagestyle{plain} 
\setcounter{page}{1}
\pagenumbering{arabic}


\section{Introduction}
\label{sec:Introduction}
Quantum chromodynamics (QCD) is the fundamental theory of the strong interaction between quarks and gluons.
One of the most important properties of QCD is that the coupling constant increases with decreasing energy.
A perturbative treatment can be applied at large momentum transfer,
while the non-perturbative effects at small momentum transfer make theoretical calculations difficult and usually need to be described by effective theories or phenomenological models with input from experiment.
The study of heavy quarkonium production in proton-proton ($pp$) collisions can provide important information to improve QCD calculations in the non-perturbative regime.
The process involves production of a $Q\overline{Q}$ pair, where $Q$ denotes a beauty or charm quark, followed by its hadronisation into a heavy quarkonium state.
In the nonrelativistic QCD (NRQCD) approach~\cite{Bodwin:1994jh,Cho:1995vh,Cho:1995ce}, both colour-singlet and colour-octet states of the intermediate $Q\overline{Q}$ pair are taken into account, and the transition probabilities from the $Q\overline{Q}$ states to the heavy quarkonium are described by non-perturbative long-distance matrix elements.
Although the NRQCD approach describes many aspects of quarkonium production successfully, no satisfactory solution has been achieved thus far to simultaneously describe differential cross-section and polarisation data over the whole kinematic region~\cite{Shao:2014yta,Zhang:2020atv}.
This puzzle can be probed via the production of heavy quarkonium pairs through the single-parton scattering (SPS) process~\cite{Sun:2014gca, Lansberg:2019fgm, Likhoded:2016zmk, Baranov:2011zz}, which can provide new tests of the NRQCD approach.

Besides SPS, heavy quarkonium pairs can be produced through double-parton scattering~(DPS)~\cite{Lansberg:2014swa}, which makes the study more complicated.
The DPS process is of great relevance since it provides valuable information on the profile and correlations of partons inside the proton.
A thorough understanding of DPS is crucial in some searches for physics beyond the standard model because several important sources of background, \eg $Z+\bbbar$, $W^\pm W^\pm$, have significant DPS contributions.
In DPS, the heavy quarkonium pair production is usually assumed to come from two independent partonic interactions, and the cross-section can be estimated according to~\cite{Calucci:1997ii,Calucci:1999yz,DelFabbro:2000ds}
\begin{equation}
\label{eq:pocket}
    \sigma^{\text{DPS}}_{\mathcal{Q}_1\mathcal{Q}_2}= \frac{1}{1+\delta_{\mathcal{Q}_1\mathcal{Q}_2}}\frac{\sigma_{\mathcal{Q}_1}\sigma_{\mathcal{Q}_2}}{\sigma_{\text{eff}}},
\end{equation}
where $\delta_{\mathcal{Q}_1\mathcal{Q}_2}$ is one if the two quarkonium states $\mathcal{Q}_1$ and $\mathcal{Q}_2$ are the same and zero if they are different,
$\sigma_{\mathcal{Q}_1}$ and $\sigma_{\mathcal{Q}_2}$ are the production cross-sections of single quarkonium $\mathcal{Q}_1$ and $\mathcal{Q}_2$ respectively, and $\sigma_{\text{eff}}$ is an effective cross-section characterising an effective transverse overlap area between the two pairs of interacting partons.

Experimentally, it is difficult to distinguish the SPS and DPS contributions.
The measurements of quarkonium pair production so far mostly separate them according to the kinematic relation between the two quarkonium decays, which is model dependent.
This paper focuses on charmonium pair production.
The production of \jpsi mesons associated with the \chic state, whose charge parity is even, through SPS is forbidden at leading order in NRQCD, resulting in a large difference in the fraction of \jpsi pairs produced from feed-down of \chic and \psitwos mesons between SPS and DPS.
The fraction of feed-down from \psitwos production is expected to be large (around 46\%) in SPS,
while it is only about 20\% in DPS~\cite{Lansberg:2014swa}.
The fraction of feed-down from \chic production is suppressed in SPS, while it is around 50\% in DPS~\cite{Lansberg:2014swa}.
Thus the feed-down fraction of \jpsi pairs from \jpsi-\psitwos or \jpsi-\chic is considered as another important test to pin down the DPS or SPS dominance.

The first evidence for \jpsi-pair production in hadron collisions was reported by the NA3 collaboration in the 1980s~\cite{NA3:1982qlq,NA3:1985rmd}.
In recent years, the \jpsi-pair production cross-section in $pp$ collisions was measured by the LHCb collaboration at ${\sqs=7\tev}$~\cite{LHCb-PAPER-2011-013} and 13\tev~\cite{LHCb-PAPER-2016-057}, by the CMS collaboration in $pp$ collisions at ${\sqs=7\tev}$~\cite{CMS:2014cmt} and by the ATLAS collaboration at ${\sqs=8\tev}$~\cite{ATLAS:2016ydt}.
It was also measured in $\proton\antiproton$ collisions by the D0 collaboration at ${\sqs=1.96\tev}$~\cite{D0:2014vql}.
However, the production of charmonium pairs involving other states, such as \jpsi-\psitwos and \jpsi-\chic has not been measured before.

In this paper, the measurement of \jpsi-\psitwos production cross-sections in $pp$ collisions at $\sqs=13\tev$ is presented, using a subset of data collected by the \lhcb experiment from 2016 to 2018 with specific trigger requirements, corresponding to an integrated luminosity of ${4.2\invfb}$.
The production cross-sections are measured for \jpsi and \psitwos mesons having transverse momentum ${\pt<14\gevc}$ and rapidity ${2.0<y<4.5}$, assuming negligible polarisation of the \jpsi and \psitwos mesons, since all the LHC measurements so far indicate that the polarisation of quarkonia is small~\cite{NRQCDpol7,NRQCDpol8,NRQCDpol9,LHCb-PAPER-2013-008,LHCb-PAPER-2013-067}.
The differential cross-sections of \jpsi-\psitwos production as functions of several kinematic variables are also measured, as well as the cross-section ratio between \jpsi-\psitwos and \jpsi-\jpsi production.

\section{Detector and dataset}
\label{sec:Detector}

The \lhcb detector~\cite{LHCb-DP-2008-001,LHCb-DP-2014-002} is a single-arm forward spectrometer covering the \mbox{pseudorapidity} range $2<\eta <5$, designed for the study of particles containing \bquark or \cquark quarks.
The detector includes a high-precision tracking system consisting of a silicon-strip vertex detector surrounding the $pp$ interaction region, a large-area silicon-strip detector located upstream of a dipole magnet with a bending power of about $4{\mathrm{\,Tm}}$, and three stations of silicon-strip detectors and straw drift tubes placed downstream of the magnet.
The tracking system provides a measurement of the momentum, \ptot, of charged particles with a relative uncertainty that varies from 0.5\% at low momentum to 1.0\% at 200\gevc.
The minimum distance of a track to a $pp$ interaction point (\ie primary vertex, PV), the impact parameter (IP), is measured with a resolution of $(15+29/\pt)\mum$, where \pt is in\,\gevc.
Different types of charged hadrons are distinguished using information from two ring-imaging Cherenkov detectors. 
Photons, electrons and hadrons are identified by a calorimeter system consisting of scintillating-pad and preshower detectors, an electromagnetic and a hadronic calorimeter.
Muons are identified by a system composed of alternating layers of iron and multiwire proportional chambers.

The \jpsi-\psitwos candidates are reconstructed using final states with two $\mumu$ pairs.
The online event selection is performed by a trigger~\cite{LHCb-DP-2012-004}, which consists of a hardware stage~(L0), based on information from the calorimeter and muon systems, followed by two stages of software trigger, HLT1 and HLT2, at which full event reconstruction is performed.
It is required that at least one $\mumu$ pair fulfils the selection criteria of the L0 and HLT1 triggers.
The L0 trigger selects two muons with the product of their transverse momentum ${{\pt}_1\times{\pt}_2>1.3^2}$, $1.5^2$ or $1.8^2\,(\text{Ge\kern -0.1em V}/c)^2$, depending on the data-taking period.
The HLT1 trigger requires two good-quality tracks with ${\pt>0.3\gevc}$ and ${p>6\gevc}$, which are loosely identified as muons, to form a \jpsi or \psitwos candidate with high invariant mass, ${m_{\mumu}>2.7\gevcc}$ or 2.9\gevcc, depending on the data-taking period.
The HLT2 trigger requires both \jpsi and \psitwos candidates to be reconstructed with good vertex-fit quality, and the invariant mass $m_{\mumu}$ is required to be within ${120\mevcc}$ of the known \jpsi or \psitwos masses~\cite{Workman:2022ynf}.

Simulated samples are produced to study the behaviour of signal candidates and determine the detection efficiencies.
The $pp$ collisions are modelled using \pythia~\cite{Sjostrand:2007gs,Sjostrand:2006za} with a specific \lhcb configuration~\cite{LHCb-PROC-2010-056}.
Decays of unstable particles are described by \evtgen~\cite{Lange:2001uf} with QED final-state radiation handled by \photos~\cite{davidson2015photos}.
The interactions of the generated particles with the detector are modelled using the \geant toolkit~\cite{Allison:2006ve,Agostinelli:2002hh} as described in Ref.~\cite{LHCb-PROC-2011-006}.
Due to limited knowledge of the associated \jpsi-\psitwos production mechanism,
the candidate selection is designed not to utilise information based on the correlation between the \jpsi and \psitwos mesons.
Therefore, the detection efficiency can be factorised into that of \jpsi and \psitwos.
The efficiency factorisation is validated using a simulated sample of ${\Upsilonres\to\jpsi\jpsi\gamma}$ decays.
The \jpsi and \psitwos efficiencies are evaluated using simulated samples of single-\jpsi and single-\psitwos candidates.
In the \pythia model, \jpsi and \psitwos mesons are generated with no polarisation, and the leading order colour-singlet and colour-octet contributions~\cite{LHCb-PROC-2010-056,Bargiotti:2007zz} are considered in prompt \jpsi and \psitwos production.

\section{Candidate selection}
\label{sec:Selection}
Further selections are performed offline to \jpsi-\psitwos candidates to reduce the combinatorial background.
Four muon tracks with good track-fit quality are selected and required to have ${1.9<\eta<4.9}$, ${\pt>0.65\gevc}$ and ${p>3\gevc}$, and to form two good-quality di-muon vertices.
The muon identification requirement is tightened to further suppress the contribution of mis-identified tracks.
The background from fake tracks is reduced by a neural-network based algorithm~\cite{DeCian:2255039}.
All events are required to have at least one reconstructed PV.
For candidates with multiple PVs in the event, the one with the smallest \chisqip is taken as the associated PV,
where \chisqip is defined as the difference in the vertex-fit $\chi^2$ of a given PV reconstructed with and without the candidate under consideration.
The four muon tracks are required to originate from the same PV.
This reduces the number of pile-up candidates, \ie~\jpsi and \psitwos candidates originating from two independent $pp$ interactions, to a negligible level.
A selection is performed on the pseudoproper time $t_z$~\cite{LHCb-PAPER-2011-003} of the \jpsi and \psitwos candidates, defined as
\begin{equation}
\label{eq:tz}
    t_z=\frac{z_{\psi}-z_{\text{PV}}}{p_z} \times m_{\psi},
\end{equation}
where $z_{\psi}$ and $z_{\text{PV}}$ are the coordinates of the \jpsi or \psitwos decay vertex and the PV along the beam axis $z$, $p_z$ is the projection of the \jpsi or \psitwos momentum along the $z$ axis, and $m_{\psi}$ is the known \jpsi or \psitwos mass~\cite{Workman:2022ynf}.
The $t_z$ uncertainty $\sigma_{t_z}$ is calculated by combining the uncertainties from $z_{\psi}$ and $z_{\text{PV}}$ since the uncertainty on $p_z$ is negligible in comparison.
Candidates with both \jpsi and \psitwos mesons having $-2<t_z<10\ps$ and $\sigma_{t_z}<0.3\ps$ are selected.
After the selection, the events with multiple candidates, in particular those in which the four muons can be combined in two different ways to form a \jpsi-\psitwos candidate, account for 1.2\% of the total. According to simulation, the candidates with wrong combination of muons follow the background distribution, so their inclusion do not introduce any bias to the signal yield determination.

\section{Cross-section determination}
\label{sec:Cross-section}
The cross-section of associated \jpsi-\psitwos production is given by
\begin{equation}
\label{eq:csDiPsi}
    \sigma_{\jpsi\text{-}\psitwos}=
    \frac{N^{\text{corr}}}{\lum\times\BR(\jpsi\to\mumu)\times\BR(\psitwos\to\mumu)},
\end{equation}
where $N^{\text{corr}}$ is the efficiency-corrected yield, \lum is the integrated luminosity, and \BR is the appropriate branching fraction.
The integrated luminosity is $\lum=4.18\pm0.08\invfb$ measured using the van der Meer scan method~\cite{LHCb:2014vhh}.
The branching fraction of the \jpsi decay is ${\BR(\jpsi\to\mumu)=(5.961\pm0.033)\%}$~\cite{Workman:2022ynf}.
Under the assumption of lepton universality in electromagnetic decays, the branching fraction of the \psitwos decay is taken to be  ${\BR(\psitwos\to\epem)=(7.93\pm0.17)\times10^{-3}}$~\cite{Workman:2022ynf}, taking advantage of its much smaller uncertainty compared to the muonic decay.

The \jpsi-\psitwos signals are extracted by performing an unbinned extended maximum likelihood fit to the two-dimensional (2D) mass distribution of \jpsi and \psitwos candidates, $(m_{\mu^+_1\mu^-_1},m_{\mu^+_2\mu^-_2})$.
The 2D mass distribution function can be factorised into the product of the two mass dimensions.
Each one-dimensional mass distribution consists of the signal and combinatorial background components.
The signal component is described by a sum of a double-sided Crystal Ball (DSCB) function~\cite{Skwarnicki:1986xj} and a Gaussian function with a common mean value but different widths.
All the tail parameters, the fraction of the DSCB function and the ratio between the two widths are fixed from simulation.
Only the common mean value and the width of the DSCB function are left as free shape parameters.
The combinatorial background distribution is modelled with an exponential function.

\begin{figure}[tb]
  \centering
    \includegraphics[width=0.49\linewidth]{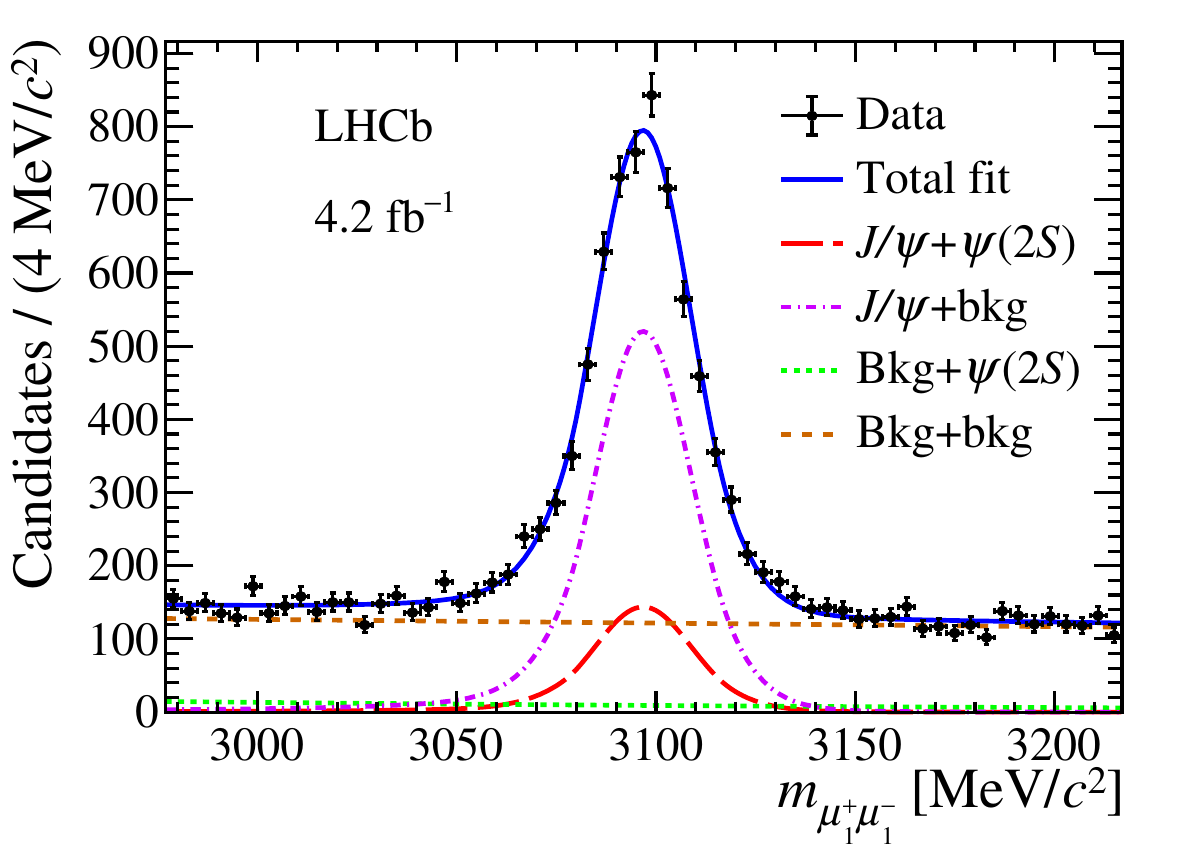}\put(-183,133){(a)}
  \hfil
    \includegraphics[width=0.49\linewidth]{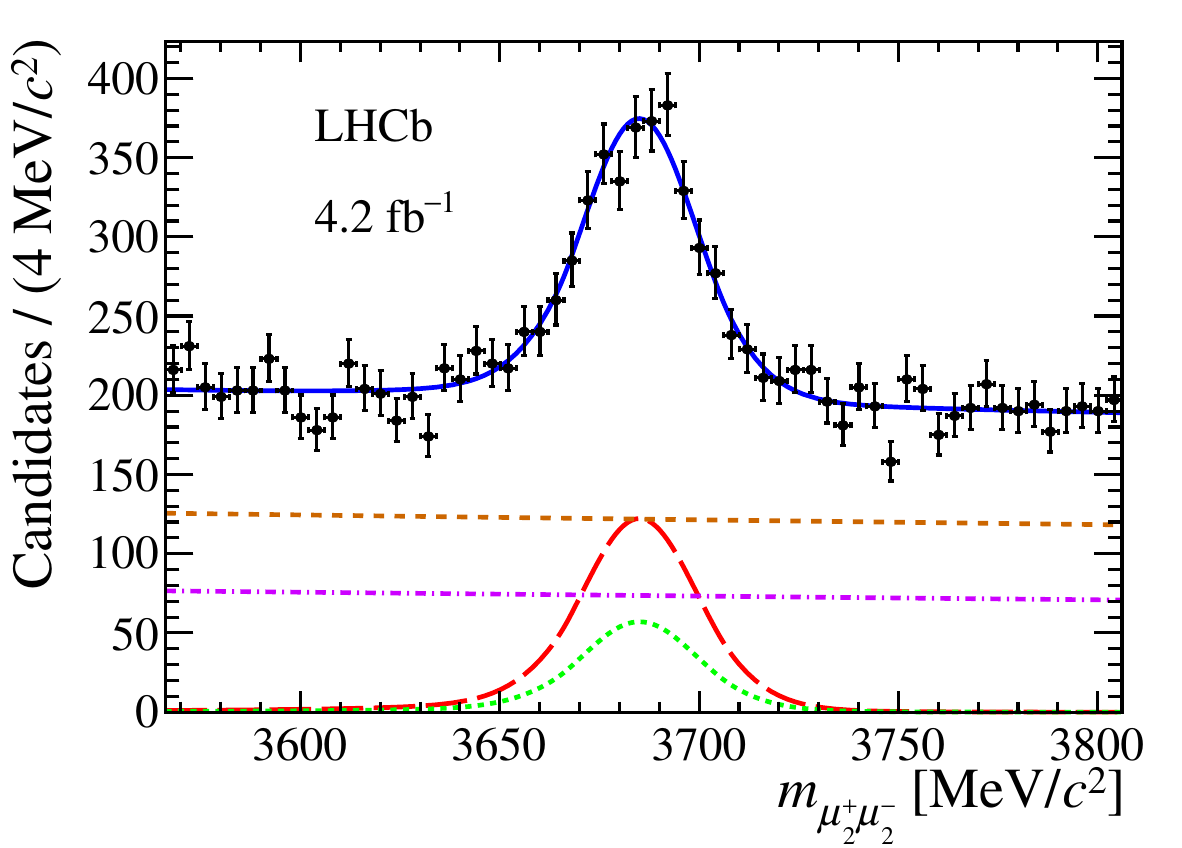}\put(-183,133){(b)}
  \caption{Distributions of (a) $m_{\mu_1^+\mu_1^-}$ and (b) $m_{\mu_2^+\mu_2^-}$ for \jpsi-\psitwos candidates together with the projections of the fit.}
  \label{fig:fitJpsiPsi2S}
\end{figure}

For \jpsi-\psitwos candidates, there are four categories:
\begin{itemize}
    \item a signal \jpsi-\psitwos decay,
    \item a true \jpsi decay with a fake \psitwos decay,
    \item a fake \jpsi decay with a true \psitwos decay,
    \item and a fake \jpsi decay with a fake \psitwos decay.
\end{itemize}
Figure~\ref{fig:fitJpsiPsi2S} shows the projections of the 2D mass distribution, together with the fit result, of the $m_{\mu^+_1\mu^-_1}$ and $m_{\mu^+_2\mu^-_2}$ invariant mass combinations.

\begin{figure}[tb]
  \centering
    \includegraphics[width=0.49\linewidth]{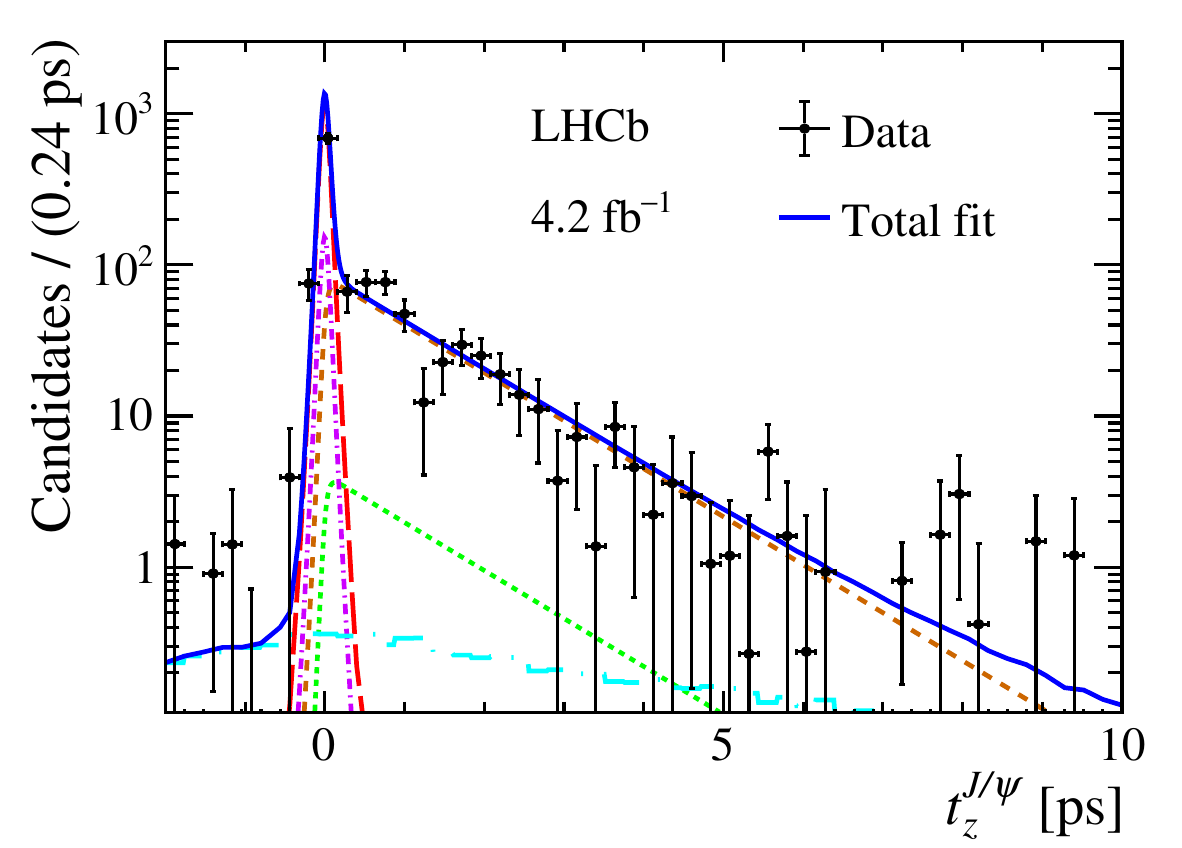}\put(-183,133){(a)}
  \hfil
    \includegraphics[width=0.49\linewidth]{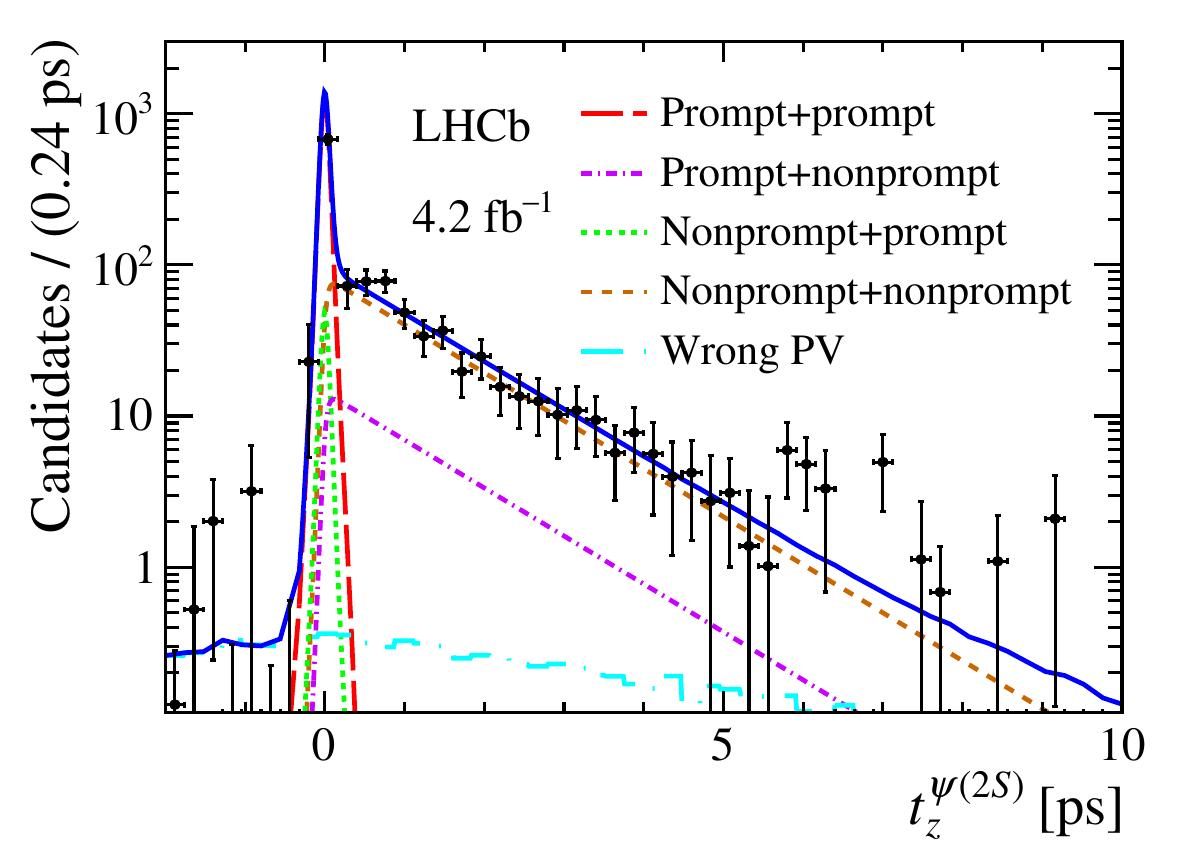}\put(-183,133){(b)}
  \caption{Distributions of (a) $t_z^{\jpsi}$ and (b) $t_z^{\psitwos}$ for \jpsi-\psitwos candidates with backgrounds subtracted together with the projections of the fit.}
  \label{fig:fitJpsiPsi2STz}
\end{figure}

In $pp$ collisions, \jpsi or \psitwos mesons can be produced either directly from hard collisions of partons, through the feed-down of excited charmonium states, or via decays of beauty hadrons.
The \jpsi or \psitwos mesons from the first two sources originate from the PV and are called prompt mesons, while those from the last source originate from decay vertices of beauty hadrons, which are typically separated from the PV, and are called nonprompt mesons.
The nonprompt contributions with the \jpsi or \psitwos meson originating from \bquark-hadron decays are subtracted by performing a maximum-likelihood fit to the 2D $t_z$ distribution of \jpsi-\psitwos candidates with backgrounds subtracted.
The background subtraction is implemented with the \sPlot method~\cite{Pivk:2004ty} using $m_{\mu_1^+\mu_1^-}$ and $m_{\mu_2^+\mu_2^-}$ as the discriminating variables.
The $t_z$ distribution of prompt \jpsi or \psitwos mesons is described by the Dirac delta function $\delta(t_z)$, while that of nonprompt mesons by an exponential function.
These distributions are convolved with a Gaussian detector resolution function.
For a very small fraction of candidates, the true PV is not reconstructed and both \jpsi and \psitwos candidates are associated to the nearest reconstructed PV.
The positions of the true PV and the wrongly associated PV are not correlated, which results in a broad distribution of $t_z$.
This component can be modelled from data using event mixing, \ie calculating $t_z$ with the \jpsi or \psitwos candidate associated to the closest PV in the next event of the sample.
Figure~\ref{fig:fitJpsiPsi2STz} shows the projections of the 2D $t_z$ distribution with backgrounds subtracted, together with the fit result, on $t_z^{\jpsi}$ and $t_z^{\psitwos}$.
The yield of prompt \jpsi-\psitwos signal candidates is $629\pm50$.

Since the kinematics of \jpsi-\psitwos decay are not known {\itshape a priori}, an efficiency correction is performed event-by-event as
\begin{equation}
\label{eq:corrN}
    N^{\text{corr}}=\sum_i \frac{w_i}{\etot_i},
\end{equation}
where the index $i$ denotes each event in data, 
$w_i$ is the \sWeight~\cite{Pivk:2004ty} corresponding to the component of prompt \jpsi-\psitwos signal candidates, 
and $\etot_i$ the detection efficiency.
The efficiency $\etot_i$ is evaluated as the product of four components:
the geometrical acceptance $\eacc_i$, the efficiency of reconstruction and selection $\erecsel_i$, the PID efficiency $\epid_i$, and the trigger efficiency $\etri_i$, giving
\begin{equation}
    \etot_i = \eacc_i\times\erecsel_i \times\epid_i\times\etri_i.
\end{equation}
For a \jpsi-\psitwos candidate, each efficiency factor can be parameterised as a function of the $\pt$ and $y$ of the \jpsi and \psitwos mesons.
As long as the real $\pt$ and $y$ values of both \jpsi and \psitwos mesons are taken, it makes no difference for the efficiencies whether the candidate is produced by SPS or DPS.
The possible kinematic correlation between two mesons, reflected in $\pt$ and $y$, carries information on the production mechanism.
On this basis, the efficiencies of each \jpsi-\psitwos candidate can be factorised into those of the \jpsi and \psitwos mesons separately, since no information related to the correlation between them is used during the reconstruction and selection process.
The efficiencies $\eacc_i$, $\erecsel_i$ and $\epid_i$ of each \jpsi-\psitwos candidate are factorised as the product of efficiencies of the \jpsi and \psitwos mesons as
\begin{equation}
    \varepsilon_i(\jpsi\text{-}\psitwos)=\varepsilon_i(\jpsi) \times\varepsilon_i(\psitwos).
\end{equation}
The trigger efficiency $\etri_i$ of each \jpsi-\psitwos candidate is factorised as
\begin{equation}
    \etri_i(\jpsi\text{-}\psitwos)= 1-\left(1-\varepsilon^{\text{L0\&HLT1}}_i(\jpsi)\right)\left(1-\varepsilon^{\text{L0\&HLT1}}_i(\psitwos)\right),
\end{equation}
where $\varepsilon^{\text{L0\&HLT1}}_i$ is the L0-and-HLT1 trigger efficiency.
The online HLT2 reconstruction algorithm is equivalent to the offline algorithm and the offline selections are tighter than the HLT2 requirements, making the HLT2 trigger fully efficient with respect to offline selected candidates.
All the efficiency terms are estimated in $(\pt,y)$ intervals of the \jpsi (\psitwos) meson using the simulated single-\jpsi (single-\psitwos) samples.
The track reconstruction efficiency, which is part of $\erecsel_i$, and the PID efficiency, $\epid_i$, are calibrated using data to eliminate differences between the simulation and data~\cite{LHCb-DP-2013-002,LHCb-DP-2013-001}.
The efficiency-corrected signal yield is determined to be $N^{\text{corr}}=(8.9\pm1.4)\times 10^3$, where the statistical uncertainty is verified by the bootstrapping approach~\cite{Efron:1979bxm}.

\section{Systematic uncertainties}
\label{sec:Systematic}

Table~\ref{table:Systematic} summaries the systematic uncertainties on the cross-section measurement.
Due to the small yields of the \jpsi-\psitwos data, 
a larger pseudo \jpsi-\psitwos sample is produced by combining the single-\jpsi and single-\psitwos candidates in data randomly
to reduce the statistical fluctuations in the evaluation of systematic uncertainties.
The same treatment of the selection, signal extraction and the efficiency correction is applied to this pseudo sample as for the \jpsi-\psitwos data sample.

An uncertainty is attributed to the imperfect modelling of the \jpsi (\psitwos) mass distribution.
As an alternative to the sum of a DSCB function and a Gaussian function,
the invariant-mass distribution of \jpsi (\psitwos) signals is described by a model derived from simulation using the kernel density estimation approach~\cite{Cranmer:2000du}.
To account for the resolution difference between data and simulation, the alternative model is convolved with a Gaussian function with a mean value of zero and the width varied freely.
The relative difference of the extracted signal yields using the default and alternative models is 1.7\%, which is taken as the systematic uncertainty.
For the measurement of differential cross-sections, this uncertainty is taken to be common to all kinematic intervals.

An uncertainty is attributed to the subtraction of nonprompt contributions. 
The quantity $\log(\chisqip)$ of the \jpsi (\psitwos) meson is used for the discrimination of nonprompt components instead of the pseudoproper time $t_z$.
The $\log(\chisqip)$ distribution of the prompt \jpsi (\psitwos) signals is described by the sum of two Bukin functions~\cite{Bukin:2007zha} with the same peak position and asymmetry, while the nonprompt component is described by one Bukin function.
The shape parameters of the Bukin functions other than the peak position and width are fixed from simulation.
The relative difference of the yield of prompt \jpsi-\psitwos signal candidates with respect to the default result, 1.9\%, is taken as the systematic uncertainty.
For the measurement of differential cross-sections, this uncertainty is considered common to all kinematic intervals.

A small fraction of \jpsi or \psitwos candidates in the \jpsi-\psitwos data may be associated to a wrong PV, which can be divided into two cases.
The first is that the true PV is reconstructed but one of the \jpsi and \psitwos candidates is associated to a wrong PV accidentally.
In this case, \jpsi and \psitwos mesons are associated to different PVs and thus are rejected by the selection.
The fraction is estimated using the simulated sample of ${\Upsilonres\to\jpsi\jpsi\gamma}$ decays to be $(0.65\pm0.02)\%$.
The second is that the true PV is not reconstructed and both \jpsi and \psitwos candidates are associated to the nearest reconstructed PV in this event.
This is a component in the 2D $t_z$ fit, with the fraction determined to be $(0.8\pm1.5)\%$.
The total fraction of wrong PV events adds up to 1.5\%, which is taken as a systematic uncertainty.
For the measurement of differential cross-sections, the fraction is studied in several intervals of \jpsi-\psitwos rapidities using the same method.
The maximum, 2.2\%, is conservatively taken as the uncertainty common to all kinematic intervals.

The limited size of the calibration samples used to determine the efficiencies can introduce a systematic uncertainty.
This is studied by using pseudoexperiments, in which the efficiency in each $(\pt,y)$ interval is sampled by Gaussian distributions with the corresponding efficiency as the mean value and the statistical uncertainty of the efficiency as the width.
The width of the cross-section distribution obtained from pseudoexperiments is determined to be 0.7\%, and taken as the systematic uncertainty.
It varies up to 3.0\% depending on the kinematic intervals for the measurement of differential cross-sections.

The binning scheme that is used to determine the efficiencies could bias the signal yield $N^{\text{corr}}$.
For the PID efficiency, this effect is studied by varying the binning schemes of the calibration sample, and the relative difference in the cross-sections is taken as an uncertainty.
For the other efficiencies, $\eacc$, $\erecsel$ and $\etri$, the kernel density estimation approach~\cite{Cranmer:2000du} is used as an alternative,
in which the efficiencies are determined as smoothed functions of $(\pt,y)$.
The relative difference in the cross-sections between the default and alternative approaches is quoted as the systematic uncertainty separately for each efficiency factor.
The uncertainty on the total efficiency is taken as a quadratic sum of these uncertainties on the four efficiency factors, and its value is 2.8\%.
For the differential cross-sections, it varies up to 5.5\% depending on the kinematic intervals.

The systematic uncertainty from the track reconstruction efficiency has two sources.
One is propagated from the statistical uncertainties of the correction factors due to the limited size of the calibration sample.
It is evaluated to be 0.9\% using pseudoexperiments.
The correction factors also depend on the event multiplicity, which is not perfectly matched between the simulation and calibration samples.
This introduces a systematic uncertainty of 0.8\% per track,
adding up to 3.2\% for the \jpsi-\psitwos signals containing four muon tracks.
The track detection efficiency uncertainty is in total 3.3\% with the two sources added in quadrature.
For the differential cross-sections, the first term varies up to 4.2\%, while the second is considered to be common to all kinematic intervals.

The systematic uncertainty associated with the trigger efficiency is evaluated by comparing the efficiency determined from the simulated single-\jpsi sample to that from the single-\jpsi data.
The trigger efficiency is determined using a subset of events that fulfil the trigger requirement with the \jpsi signals excluded~\cite{LHCb-PUB-2014-039}.
Due to the small yields of the \psitwos sample, the ratio of trigger efficiencies between data and simulation determined using the \jpsi events is also applied to \psitwos mesons.
The relative difference in the \jpsi-\psitwos cross-section between using the trigger efficiency obtained from data and simulation is 0.7\%, and is taken as a systematic uncertainty.
For the differential cross-sections, the quoted uncertainty varies up to 7.5\% depending on the kinematic intervals.

The uncertainties on the branching fractions lead to an uncertainty of 2.2\% on the measured cross-section.
The luminosity is determined with a relative uncertainty of 2.0\% using the methods described in Ref.~\cite{LHCb:2014vhh}.
With these uncertainties due to independent effects added in quadrature, the total systematic uncertainty on the \jpsi-\psitwos production cross-section is determined to be 5.9\%.

\begin{table}[tb]
    \centering
    \caption{Summary of relative systematic uncertainties on the measurement of \jpsi-\psitwos production cross-section. The total systematic uncertainty is the quadratic sum of the individual contributions.}
    \label{table:Systematic}
    \begin{tabular}{lc}
	\hline
	Source                       		& Uncertainty    \\
	\hline
	Signal mass model               	& 1.7\% 			\\
	Nonprompt contribution              & 1.9\% 			\\
	Wrong PV association         		& 1.5\% 			\\
	Calibration sample statistics       & 0.7\% 			\\
	Efficiency determination            & 2.1\% 			\\
	Track reconstruction efficiency     & 3.3\% 			\\
	Trigger efficiency                  & 0.7\% 			\\
	Branching fractions           		& 2.2\% 			\\
	Luminosity                   		& 2.0\% 			\\ 
	\hline
	Total                               & 5.9\% 			\\
	\hline
    \end{tabular}
\end{table}

\section{Production cross-section and cross-section ratio}
\label{sec:Production}
The cross-section of associated \jpsi-\psitwos production with both \jpsi and \psitwos mesons having ${\pt<14\gevc}$ and ${2.0<y<4.5}$ is measured to be
\begin{equation*}
    \sigma_{\jpsi\text{-}\psitwos}= 4.5\pm0.7~(\text{stat})\pm0.3~(\text{syst})\nb,
\end{equation*}
under the assumption of negligible polarisation of the \jpsi and \psitwos mesons.
The detection efficiency of \jpsi and \psitwos mesons depends on their polarisation, especially on the polarisation parameter $\lambda_{\theta}$~\cite{LHCb-PAPER-2013-008,LHCB-PAPER-2015-037,LHCb-PAPER-2013-067,LHCb-PAPER-2018-049}.
If the polarisation parameter $\lambda_{\theta}$ is assumed to be $+0.2$ ($-0.2$) for both \jpsi and \psitwos mesons in the helicity frame, the cross-section of \jpsi-\psitwos production changes by $+6.5\%$ ($-6.6\%$).

According to Eq.~\ref{eq:pocket}, the ratio between the product of single-\jpsi and single-\psitwos cross-sections and the \jpsi-\psitwos cross-section, $\cfrac{\sigma_{\jpsi}\sigma_{\psitwos}}{\sigma_{\jpsi\text{-}\psitwos}}$
is a lower limit of $\sigma_{\text{eff}}$.
It can be interpreted as the effective cross-section $\sigma_{\text{eff}}$ if the \mbox{\jpsi-\psitwos} candidates are produced entirely from the DPS process.
The cross-section ratio is determined to be
\begin{equation*}
    \frac{\sigma_{\jpsi}\sigma_{\psitwos}}{\sigma_{\jpsi\text{-}\psitwos}}=7.1\pm1.1~(\text{stat})\pm0.8~(\text{syst})\mbarn,
\end{equation*}
where the single-\jpsi cross-section in $pp$ collisions at $\sqs=13\tev$ with $\pt<14\gevc$ and $2.0<y<4.5$ is ${\sigma_{\jpsi}=15.03\pm0.94\mub}$~\cite{LHCB-PAPER-2015-037}.
The production cross-section of \psitwos mesons in the range of $2<\pt<14\gevc$ and $2.0<y<4.5$ in $pp$ collisions at $\sqs=13\tev$ is measured to be ${1.42\pm0.10\mub}$ by LHCb~\cite{LHCb-PAPER-2018-049}.
The \psitwos cross-section with $\pt<2\gevc$ and $2.0<y<4.5$ is estimated
to be ${0.70\pm0.11\mub}$~\cite{LHCb-PAPER-2011-045,LHCB-PAPER-2015-037} assuming that the production ratio of \jpsi and \psitwos does not depend on the collision energy~\cite{PHENIX:2016vmz}.
Combining these, the single-\psitwos cross-section in the fiducial range $\pt<14\gevc$ and $2.0<y<4.5$ is estimated to be ${\sigma_{\psitwos}=2.12\pm0.15\mub}$.
The systematic uncertainties of the single-\jpsi, single-\psitwos and \jpsi-\psitwos cross-sections are treated as uncorrelated.

\subsection{Differential cross-sections}
\label{sec:diffJpsiPsi2S}
The differential cross-section of \jpsi-\psitwos production is measured as a function of the absolute value of the rapidity difference between the \jpsi and \psitwos mesons $\Delta y$, the absolute value of the difference in the azimuthal angle $\phi$ between the \jpsi and \psitwos mesons $\Delta\phi$, the transverse momentum $\pt^{\jpsi\text{-}\psitwos}$, rapidity $y_{\jpsi\text{-}\psitwos}$ and invariant mass $m_{\jpsi\text{-}\psitwos}$ of the \jpsi-\psitwos candidates.
The binning scheme is chosen to have adequate and approximately equal signal yields in each bin.
Due to the limited signal yields, the differential cross-sections of \jpsi-\psitwos production are calculated using the \sWeights based on the global mass fit and global $t_z$ fit rather than from the fit in each kinematic interval.
The potential correlation between these kinematic variables and the discriminating variables used in the \sPlot method may bias the signal yields.
Correction factors are thus obtained using a pseudo \jpsi-\psitwos sample produced by combining single-\jpsi and single-\psitwos candidates in data randomly.
They are calculated as the ratio of yields obtained from the fit in each kinematic interval and from the global \sWeights.
The uncertainties on the correction factors come from the limited size of the pseudosample, resulting in another systematic uncertainty on the differential cross-sections of \jpsi-\psitwos production.
It varies up to 10.7\% depending on the kinematic intervals.
The other systematic uncertainties on differential cross-sections are presented in Section~\ref{sec:Systematic}.
The measured differential cross-sections of \jpsi-\psitwos production are shown in Figure~\ref{fig:comTheoryJpsiPsi2S}, and listed in Tables~\ref{table:csJpsiPsi2SDY}--\ref{table:csJpsiPsi2SM} in Appendix~\ref{sec:Tables}.

\begin{figure}[tb]
  \centering
    \includegraphics[width=0.45\linewidth]{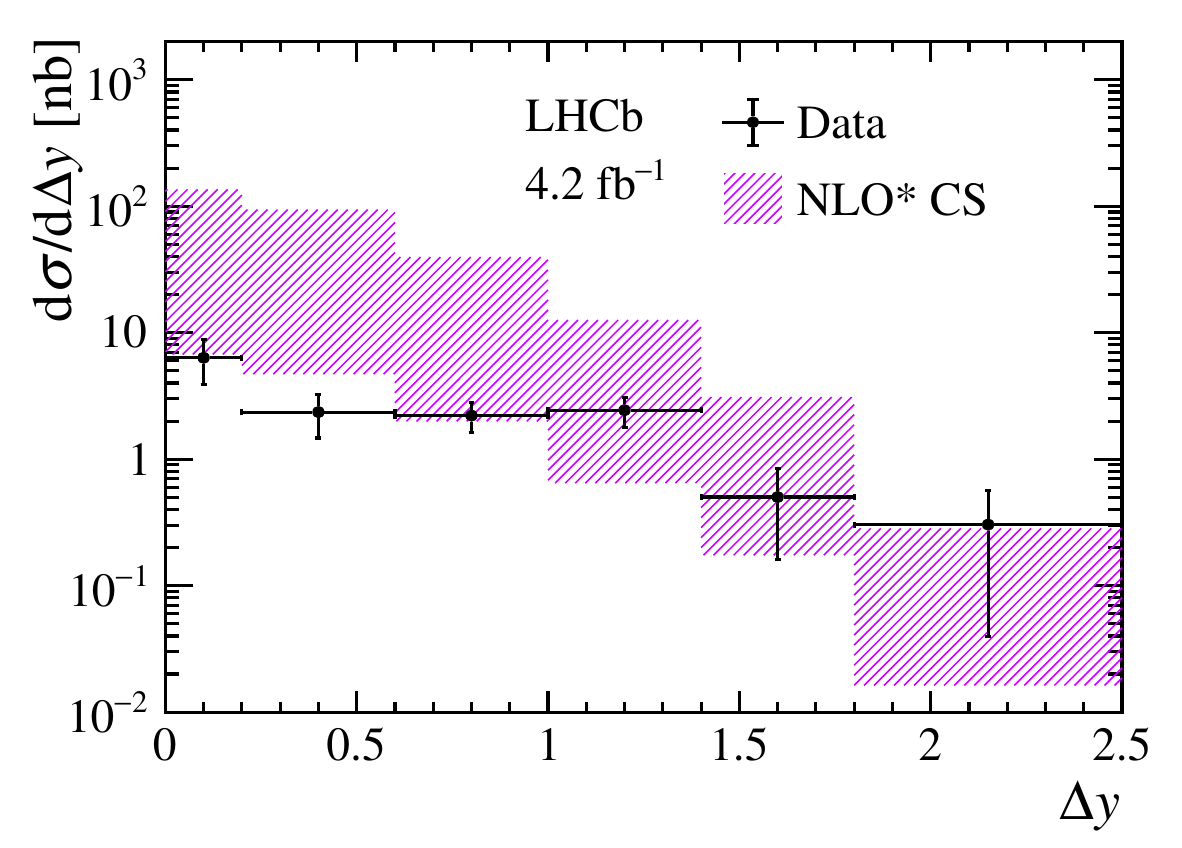}\put(-165,120){(a)}
    \includegraphics[width=0.45\linewidth]{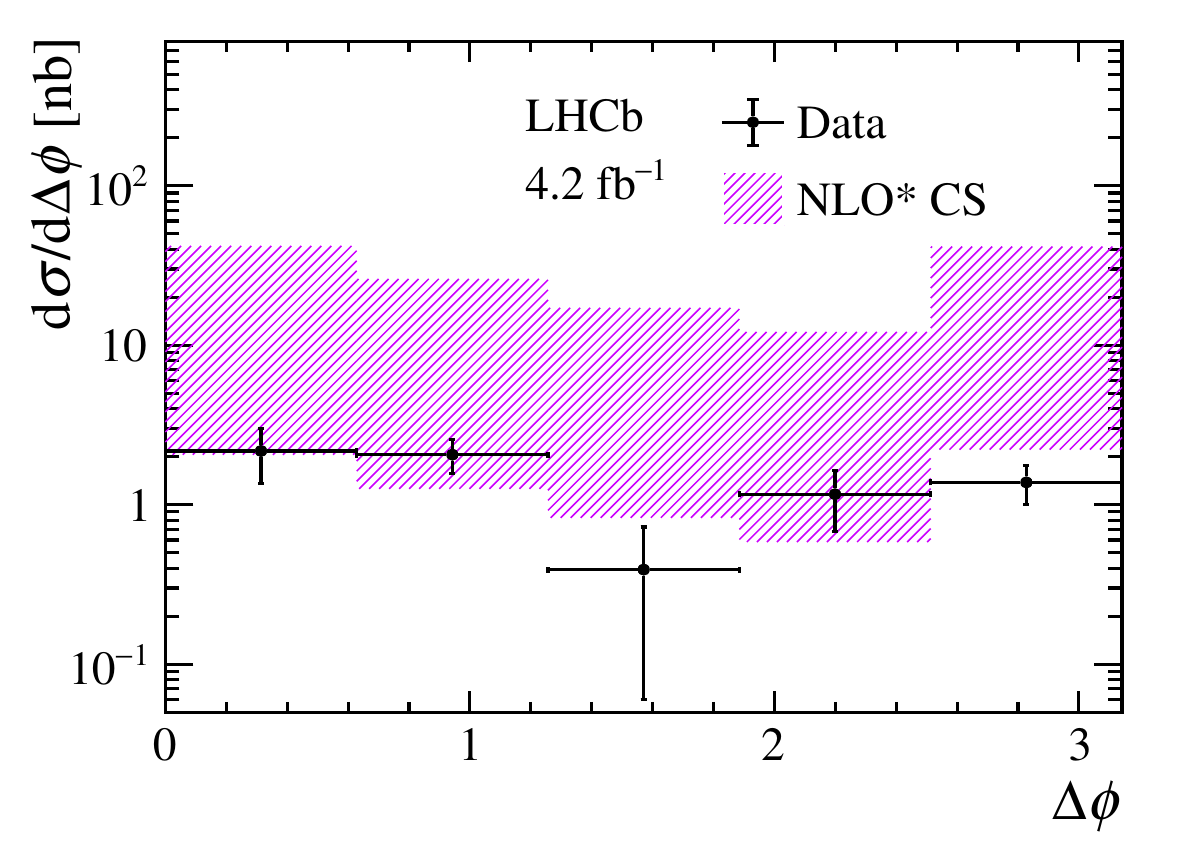}\put(-165,120){(b)}
  \\
    \includegraphics[width=0.45\linewidth]{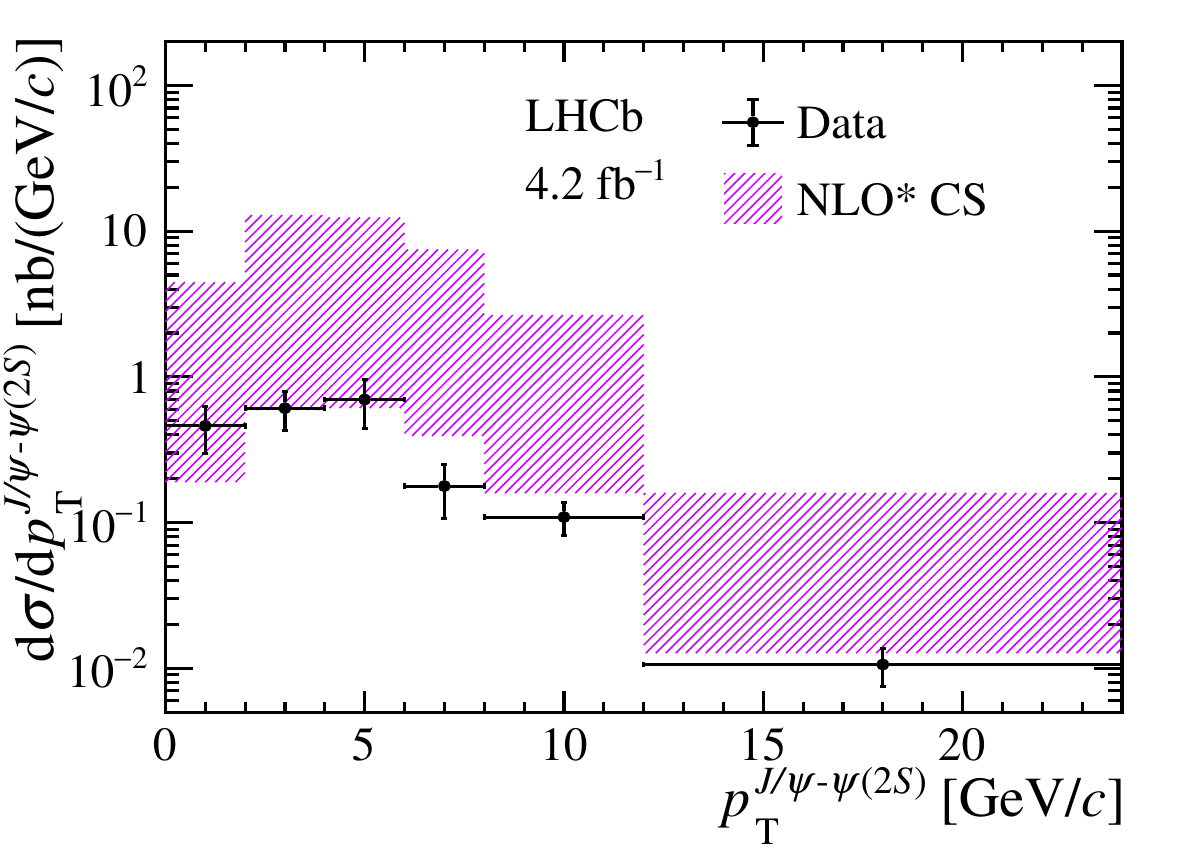}\put(-165,120){(c)}
    \includegraphics[width=0.45\linewidth]{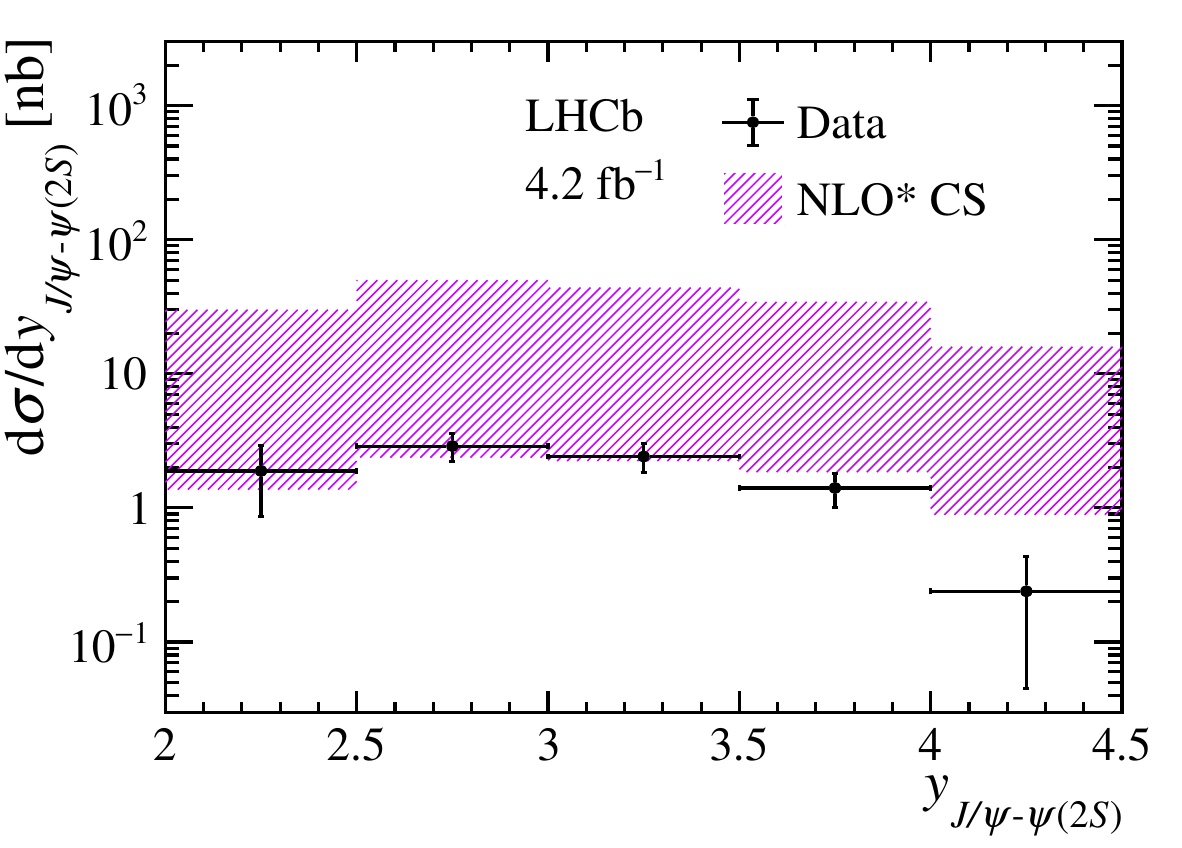}\put(-165,120){(d)}
  \\
    \includegraphics[width=0.45\linewidth]{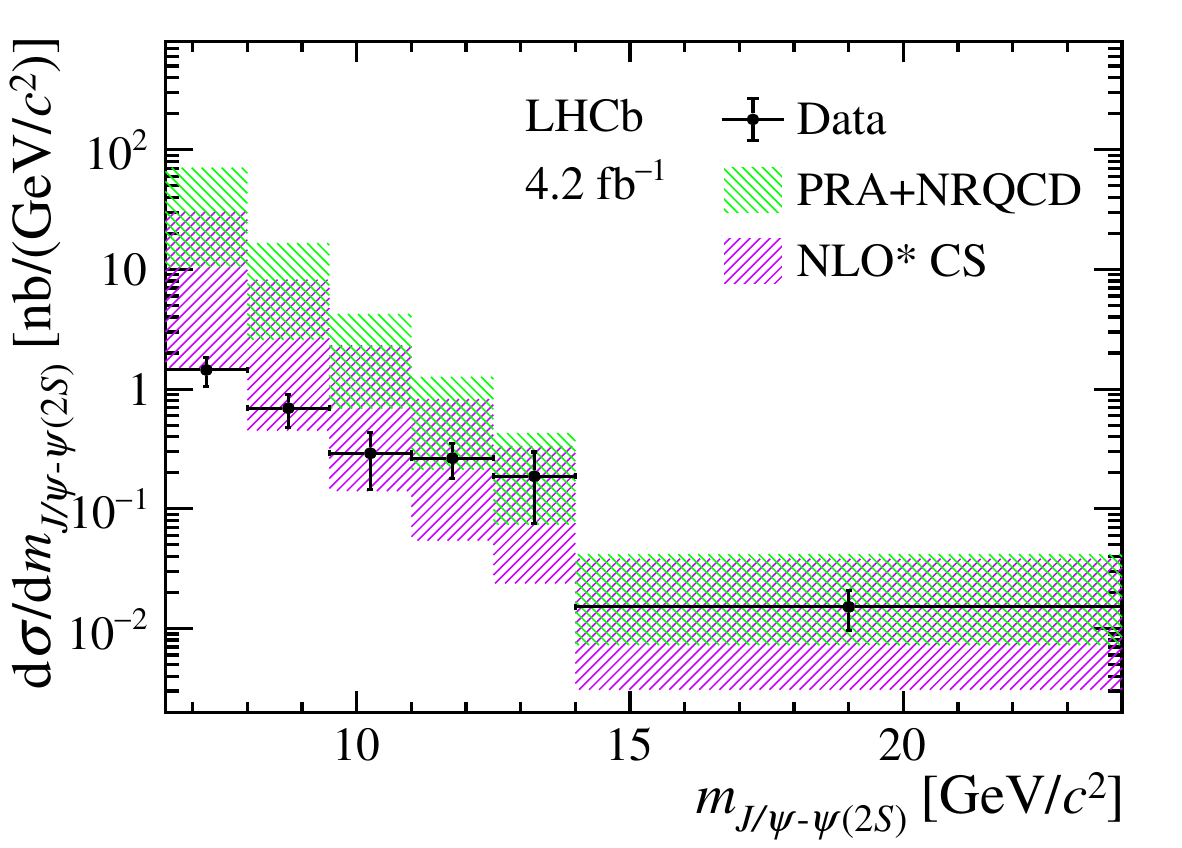}\put(-165,120){(e)}
  \caption{Differential cross-section of \jpsi-\psitwos production as a function of (a) $\Delta y$, (b) $\Delta\phi$, (c) $\pt^{\jpsi\text{-}\psitwos}$, (d) $y_{\jpsi\text{-}\psitwos}$ and (e) $m_{\jpsi\text{-}\psitwos}$, compared with the NLO*~CS predictions for SPS~\cite{Lansberg:2013qka,Shao:2012iz,Shao:2015vga}. The $m_{\jpsi\text{-}\psitwos}$ spectrum is also compared with the PRA+NRQCD predictions for SPS~\cite{He:2019qqr}. The data contain contributions from both SPS and DPS.}
  \label{fig:comTheoryJpsiPsi2S}
\end{figure}
In the NRQCD approach, explicit calculations~\cite{Lansberg:2019fgm} indicate the colour-octet contributions are much smaller than colour-singlet contributions.
The incomplete (not including loop diagrams) next-to-leading order colour-singlet (NLO*~CS) predictions~\cite{Lansberg:2013qka} of \jpsi-\psitwos production in the SPS process can be obtained from HELAC-Onia~\cite{Shao:2012iz,Shao:2015vga}, an automatic matrix element generator for heavy quarkonium physics.
The NLO*~CS predictions are compared with the measured \jpsi-\psitwos production cross-sections as shown in Figure~\ref{fig:comTheoryJpsiPsi2S}.
The theoretical uncertainties include the uncertainty due to factorisation and renormalisation scales, which is dominant, and the PDF uncertainty.
The data and theory are consistent within the large theoretical uncertainties,
albeit the DPS contribution is not subtracted from the measurements.

As shown in Figure~\ref{fig:comTheoryJpsiPsi2S}(e), the $m_{\jpsi\text{-}\psitwos}$ spectrum is also compared with the SPS predictions combining the parton Reggeization approach (PRA)~\cite{Karpishkov:2017kph} and the NRQCD factorisation approach~\cite{He:2019qqr}, which includes a subset of higher-order QCD corrections in an infrared-safe way without ad-hoc kinematic cuts.
Only the scale uncertainties are considered in the predictions.
The PRA+NRQCD predictions for SPS are larger than the SPS+DPS data at small $m_{\jpsi\text{-}\psitwos}$, and consistent with them at large $m_{\jpsi\text{-}\psitwos}$.

\subsection{Cross-section ratio between \boldmath \jpsi-\psitwos and \jpsi-\jpsi production}
\begin{figure}[tb]
  \centering
    \includegraphics[width=0.49\linewidth]{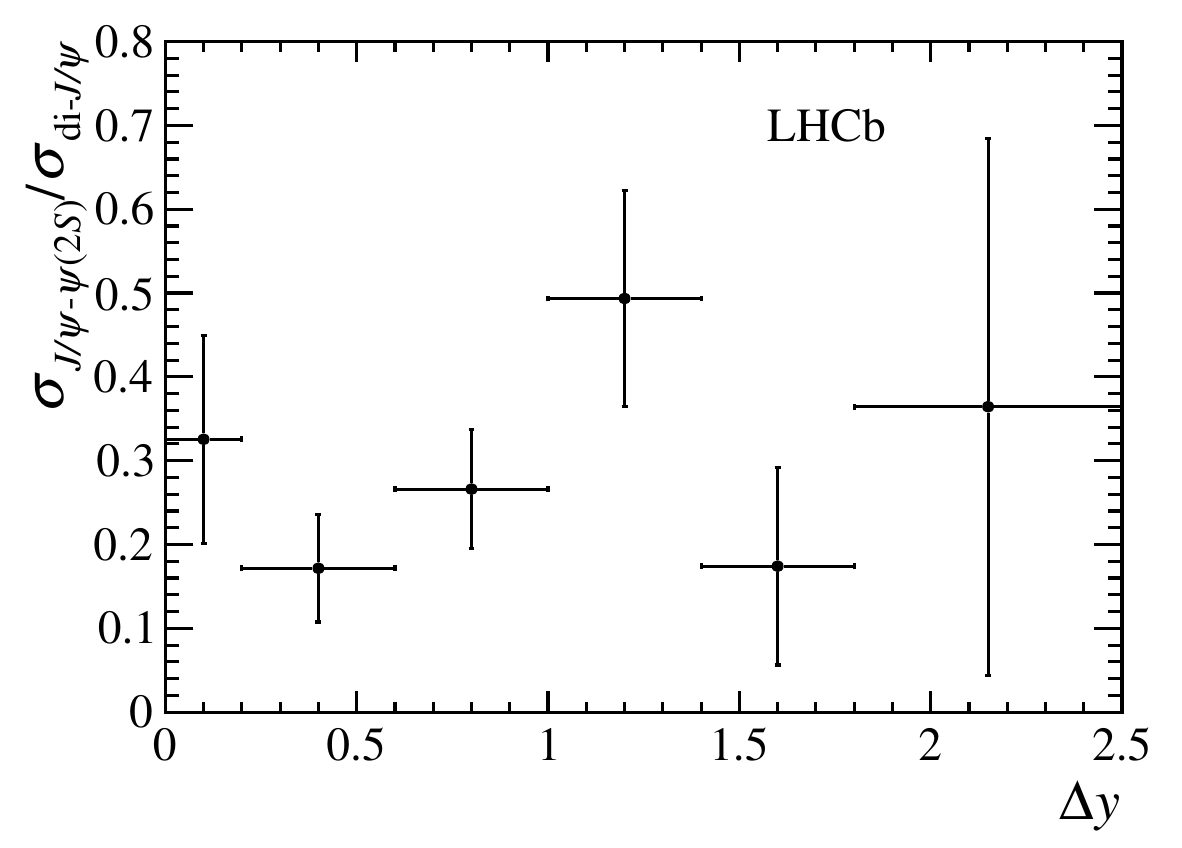}\put(-183,133){(a)}
  \hfil
    \includegraphics[width=0.49\linewidth]{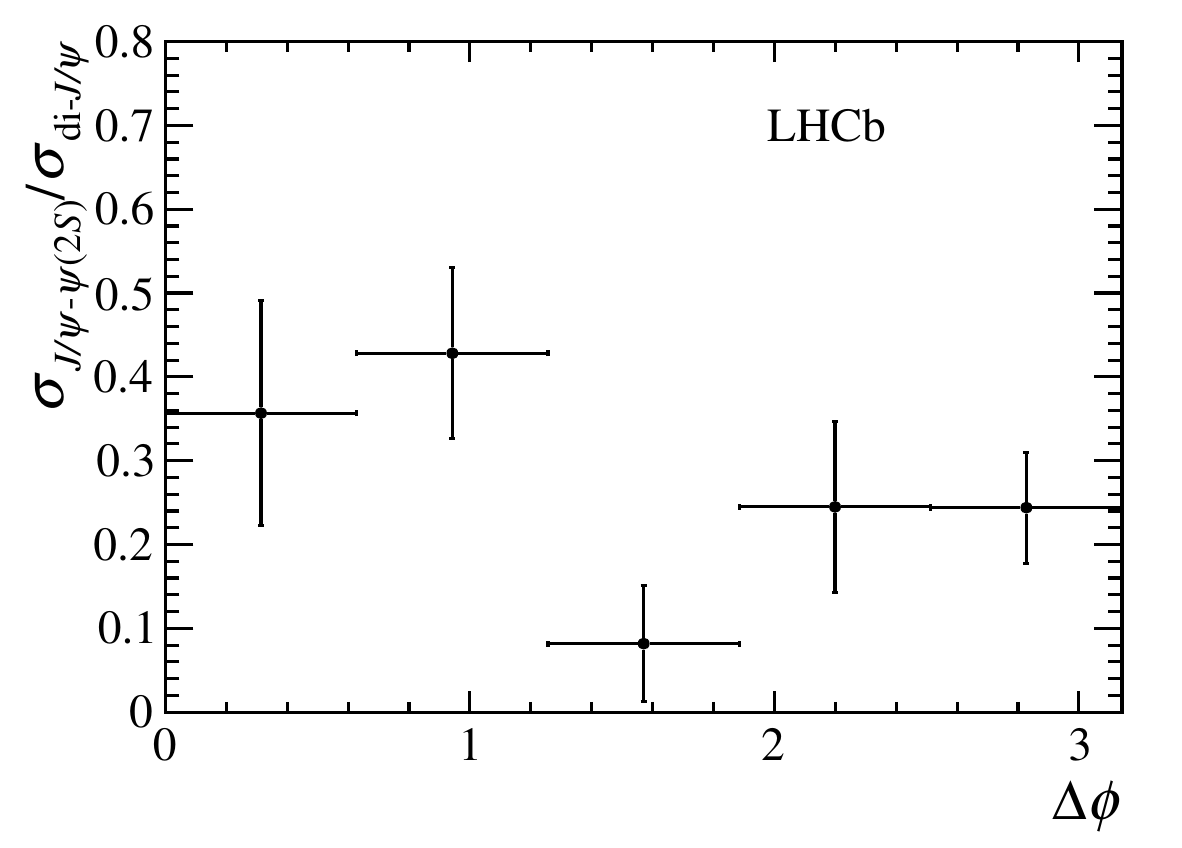}\put(-183,133){(b)}
  \\
    \includegraphics[width=0.49\linewidth]{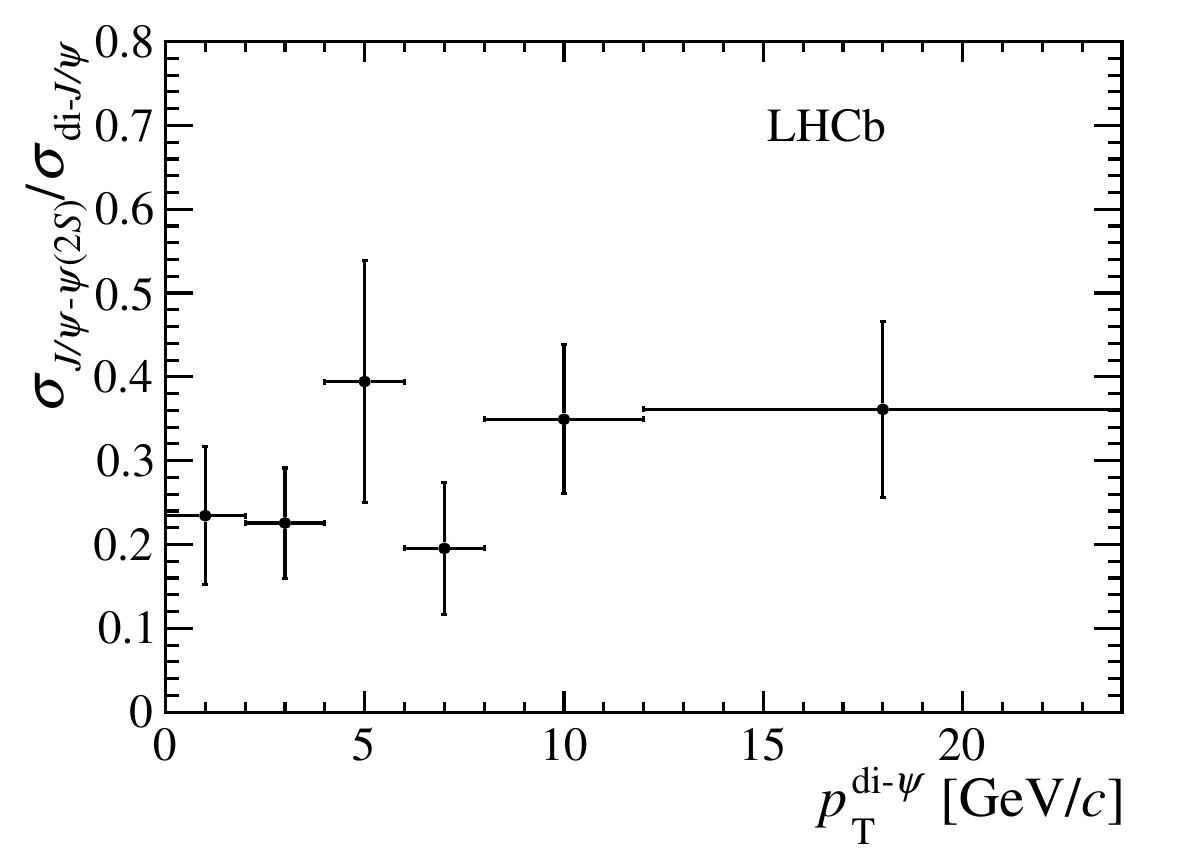}\put(-183,133){(c)}
  \hfil
    \includegraphics[width=0.49\linewidth]{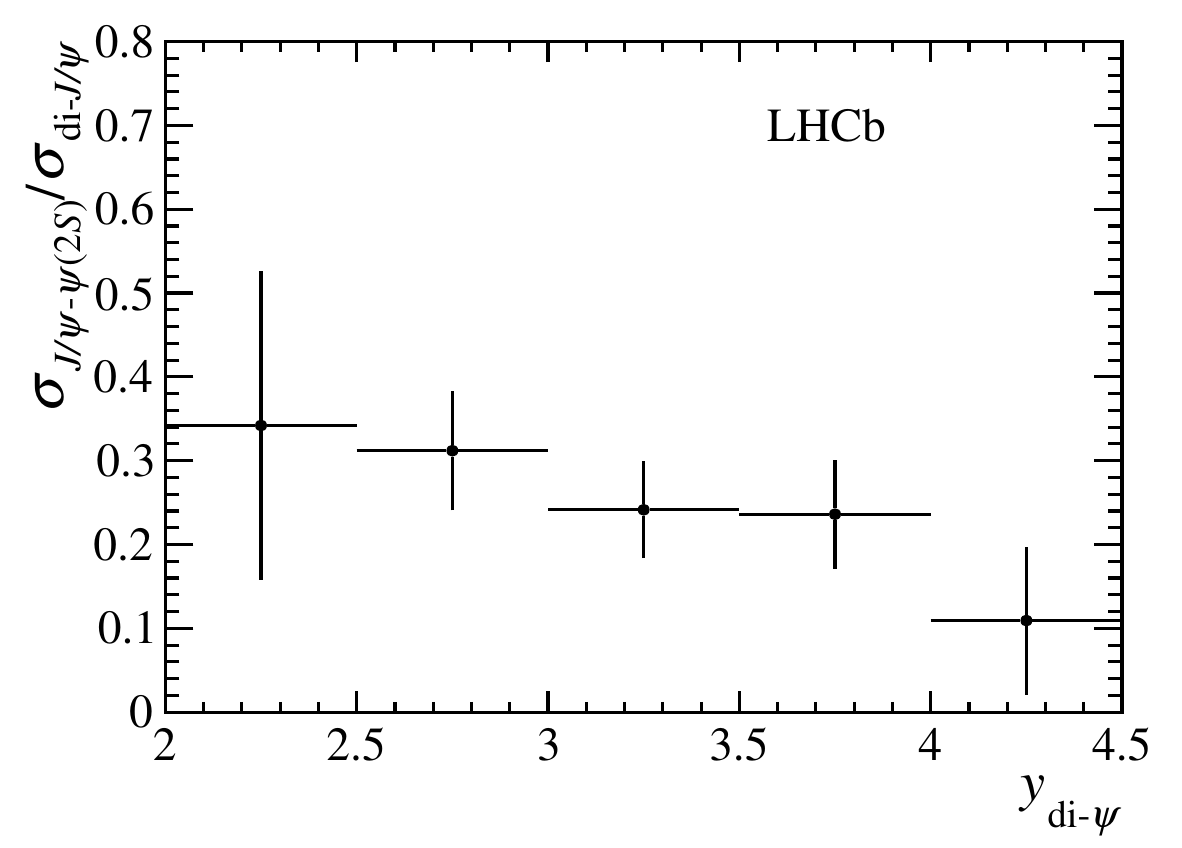}\put(-183,133){(d)}
  \caption{Cross-section ratios between \jpsi-\psitwos and \jpsi-\jpsi production as a function of (a) $\Delta y$, (b) $\Delta\phi$, (c) $\pt^{\text{di-}\psi}$ and (d) $y_{\text{di-}\psi}$. The error bars represent the statistical and systematic uncertainties added in quadrature. The symbol di-$\psi$ indicates \jpsi-\jpsi or \jpsi-\psitwos.}
  \label{fig:csRatio}
\end{figure}
The cross-section ratio between \jpsi-\psitwos and \jpsi-\jpsi production is measured to be
\begin{equation*}
    \frac{\sigma_{\jpsi\text{-}\psitwos}}{\sigma_{\jpsi\text{-}\jpsi}}=0.274\pm0.044~(\text{stat})\pm0.008~(\text{syst}),
\end{equation*}
in which the \jpsi-\jpsi production measurement is performed under a similar strategy as this analysis, as described in Ref.~\cite{LHCb-PAPER-2023-022}.
The systematic uncertainties due to the signal mass model, nonprompt contribution, wrong PV association, efficiency determination, multiplicity dependence of the track detection efficiency, trigger efficiency and luminosity measurement are thus assumed to be fully correlated.
The uncertainties from the branching fractions cancel partially and the remaining uncertainty is 2.2\% on the cross-section ratio.
The uncertainties due to calibration sample sizes are considered uncorrelated.
The statistical uncertainty on the cross-section ratio is dominant.
The ratio is predicted to be $0.94\pm0.30$ for SPS~\cite{Lansberg:2014swa}, 
and $0.282\pm0.027$ for DPS, according to Eq.~\ref{eq:pocket} and the measured single-\jpsi and single-\psitwos cross-sections~\cite{LHCb-PAPER-2011-045,LHCB-PAPER-2015-037,LHCb-PAPER-2018-049} assuming that the $\sigma_{\text{eff}}$ is the same for these two processes.
The measured value is much smaller than the SPS prediction and consistent with the DPS prediction, indicating a prominent DPS contribution to charmonium pair production.

The cross-section ratio is also measured as a function of $\Delta y$, $\Delta\phi$, $\pt^{\text{di-}\psi}$ and $y_{\text{di-}\psi}$, as shown in Figure~\ref{fig:csRatio}.
The symbol di-$\psi$ indicates \jpsi-\jpsi or \jpsi-\psitwos.
The results are also presented in Tables~\ref{table:csRatioDY}--\ref{table:csRatioY} in Appendix~\ref{sec:Tables}.
No clear dependence of the cross-section ratio on the kinematic variables is seen given the large uncertainties so far.

\section{Conclusion}
\label{sec:Conclusion}
The cross-section of associated \jpsi-\psitwos production in $pp$ collisions at $\sqs=13\tev$ is measured to be $4.5\pm0.7~(\text{stat})\pm0.3~(\text{syst})\nb$ for \jpsi and \psitwos mesons with ${\pt<14\gevc}$ and ${2.0<y<4.5}$, using a data sample corresponding to an integrated luminosity of $4.2\invfb$ collected by the LHCb experiment.
The differential cross-sections of \jpsi-\psitwos production are measured as functions of $\Delta y$, $\Delta\phi$, $\pt^{\jpsi\text{-}\psitwos}$, $y_{\jpsi\text{-}\psitwos}$ and $m_{\jpsi\text{-}\psitwos}$, 
and compared to the NLO*~CS predictions for the SPS process.
The obtained results are consistent with the theory given the presence of large theoretical uncertainties, although the DPS contribution is not subtracted from the measurement.
The PRA+NRQCD predictions for SPS are larger than the data at small $m_{\jpsi\text{-}\psitwos}$ and consistent with them at large $m_{\jpsi\text{-}\psitwos}$.
The cross-section ratio between \jpsi-\psitwos and \jpsi-\jpsi production is also measured, 
and found to be significantly smaller than the SPS prediction and consistent with the DPS expectation.
Previously~\cite{LHCb-PAPER-2016-057,CMS:2014cmt,ATLAS:2016ydt,D0:2014vql}, a prominent DPS contribution to the charmonium pair production was established mainly from the study of the kinematic relation of the two charmonia states. This is confirmed in this paper in a novel way.

\section*{Acknowledgements}
%
%
\noindent We would like to thank H.-S. Shao, J.-P. Lansberg, Z.-G. He and M. A. Nefedov for the helpful inputs on theoretical calculations of the \jpsi-\psitwos production.
We express our gratitude to our colleagues in the CERN
accelerator departments for the excellent performance of the LHC. We
thank the technical and administrative staff at the LHCb
institutes.
We acknowledge support from CERN and from the national agencies:
CAPES, CNPq, FAPERJ and FINEP (Brazil); 
MOST and NSFC (China); 
CNRS/IN2P3 (France); 
BMBF, DFG and MPG (Germany); 
INFN (Italy); 
NWO (Netherlands); 
MNiSW and NCN (Poland); 
MCID/IFA (Romania); 
MICINN (Spain); 
SNSF and SER (Switzerland); 
NASU (Ukraine); 
STFC (United Kingdom); 
DOE NP and NSF (USA).
We acknowledge the computing resources that are provided by CERN, IN2P3
(France), KIT and DESY (Germany), INFN (Italy), SURF (Netherlands),
PIC (Spain), GridPP (United Kingdom), 
CSCS (Switzerland), IFIN-HH (Romania), CBPF (Brazil),
and Polish WLCG (Poland).
We are indebted to the communities behind the multiple open-source
software packages on which we depend.
Individual groups or members have received support from
ARC and ARDC (Australia);
Key Research Program of Frontier Sciences of CAS, CAS PIFI, CAS CCEPP, 
Fundamental Research Funds for the Central Universities, 
and Sci. \& Tech. Program of Guangzhou (China);
Minciencias (Colombia);
EPLANET, Marie Sk\l{}odowska-Curie Actions, ERC and NextGenerationEU (European Union);
A*MIDEX, ANR, IPhU and Labex P2IO, and R\'{e}gion Auvergne-Rh\^{o}ne-Alpes (France);
AvH Foundation (Germany);
ICSC (Italy); 
GVA, XuntaGal, GENCAT, Inditex, InTalent and Prog.~Atracci\'on Talento, CM (Spain);
SRC (Sweden);
the Leverhulme Trust, the Royal Society
 and UKRI (United Kingdom).


\newcommand{\xx}{\ensuremath{\kern 0.5em }}
\newcommand{\xxx}{\ensuremath{\kern 0.75em }}
\clearpage

\appendix

\section{Differential cross-sections and cross-section ratios}
\label{sec:Tables}

The measured differential cross-sections of \jpsi-\psitwos production as functions of $\Delta y$, $\Delta\phi$, $\pt^{\jpsi\text{-}\psitwos}$, $y_{\jpsi\text{-}\psitwos}$ and $m_{\jpsi\text{-}\psitwos}$ are listed in Tables~\ref{table:csJpsiPsi2SDY}--\ref{table:csJpsiPsi2SM}.
The cross-section ratios between \jpsi-\psitwos and \jpsi-\jpsi production as functions of $\Delta y$, $\Delta\phi$, $\pt^{\text{di-}\psi}$ and $y_{\text{di-}\psi}$ are listed in Tables~\ref{table:csRatioDY}--\ref{table:csRatioY}.

\begin{table}[h]
\centering
\caption{Differential cross-section $\deriv\sigma/\deriv\Delta y$ [nb] of \jpsi-\psitwos production. The first uncertainty is statistical and the second systematic.}
\label{table:csJpsiPsi2SDY}
\begin{tabular}{cc}
\hline $\Delta y$ & $\deriv\sigma/\deriv\Delta y$ [nb] \\\hline
0.0--0.2& $ 6.33\pm 2.37\pm 0.55$ \\
0.2--0.6& $ 2.36\pm 0.87\pm 0.15$ \\
0.6--1.0& $ 2.21\pm 0.58\pm 0.15$ \\
1.0--1.4& $ 2.44\pm 0.62\pm 0.19$ \\
1.4--1.8& $ 0.50\pm 0.34\pm 0.05$ \\
1.8--2.5& $ 0.30\pm 0.26\pm 0.04$ \\
\hline
\end{tabular}
\end{table}

\begin{table}[h]
\centering
\caption{Differential cross-section $\deriv\sigma/\deriv\Delta\phi$ [nb] of \jpsi-\psitwos production. The first uncertainty is statistical and the second systematic.}
\label{table:csJpsiPsi2SDPHI}
\begin{tabular}{cc}
\hline $\Delta\phi/\pi$ & $\deriv\sigma/\deriv\Delta\phi$ [nb] \\\hline
0.0--0.2& $16.31\pm 0.80\pm 1.22$ \\
0.2--0.4& $15.47\pm 0.47\pm 1.12$ \\
0.4--0.6& $\xx2.94\pm 0.33\pm 0.24$ \\
0.6--0.8& $\xx8.73\pm 0.48\pm 0.64$ \\
0.8--1.0& $10.36\pm 0.36\pm 0.76$ \\
\hline
\end{tabular}
\end{table}

\begin{table}[h]
\centering
\caption{Differential cross-section $\deriv\sigma/\deriv\pt^{\jpsi\text{-}\psitwos}$ [nb/(\ensuremath{\text{Ge\kern -0.1em V}/c})] of \jpsi-\psitwos production. The first uncertainty is statistical and the second systematic.}
\label{table:csJpsiPsi2SPT}
\begin{tabular}{cc}
\hline $\pt^{\jpsi\text{-}\psitwos}$ [\ensuremath{\text{Ge\kern -0.1em V}/c}] & $\deriv\sigma/\deriv \pt^{\jpsi\text{-}\psitwos}$ [nb/(\ensuremath{\text{Ge\kern -0.1em V}/c})] \\\hline
0--2& $ 0.461\pm 0.158\pm 0.039$ \\
2--4& $ 0.609\pm 0.174\pm 0.045$ \\
4--6& $ 0.699\pm 0.252\pm 0.049$ \\
6--8& $ 0.178\pm 0.071\pm 0.013$ \\
8--12& $ 0.109\pm 0.027\pm 0.008$ \\
12--24& $ 0.011\pm 0.003\pm 0.001$ \\
\hline
\end{tabular}
\end{table}

\begin{table}[]
\centering
\caption{Differential cross-section $\deriv\sigma/\deriv y_{\jpsi\text{-}\psitwos}$ [nb] of \jpsi-\psitwos production. The first uncertainty is statistical and the second systematic.}
\label{table:csJpsiPsi2SY}
\begin{tabular}{cc}
\hline $y_{\jpsi\text{-}\psitwos}$ & $\deriv\sigma/\deriv y_{\jpsi\text{-}\psitwos}$ [nb] \\\hline
2.0--2.5& $ 1.89\pm 0.99\pm 0.22$ \\
2.5--3.0& $ 2.89\pm 0.64\pm 0.21$ \\
3.0--3.5& $ 2.42\pm 0.57\pm 0.16$ \\
3.5--4.0& $ 1.41\pm 0.38\pm 0.12$ \\
4.0--4.5& $ 0.24\pm 0.19\pm 0.04$ \\
\hline
\end{tabular}
\end{table}

\begin{table}[]
\centering
\caption{Differential cross-section $\deriv\sigma/\deriv m_{\jpsi\text{-}\psitwos}$ [nb/(\ensuremath{\text{Ge\kern -0.1em V}/c^2})] of \jpsi-\psitwos production. The first uncertainty is statistical and the second systematic.}
\label{table:csJpsiPsi2SM}
\begin{tabular}{cc}
\hline $m_{\jpsi\text{-}\psitwos}$ [\ensuremath{\text{Ge\kern -0.1em V}/c^2}] & $\deriv\sigma/\deriv m_{\jpsi\text{-}\psitwos}$ [nb/(\ensuremath{\text{Ge\kern -0.1em V}/c^2})] \\\hline
6.5--8.0& $ 1.446\pm 0.378\pm 0.101$ \\
8.0--9.5& $ 0.693\pm 0.207\pm 0.047$ \\
9.5--11.0& $ 0.291\pm 0.144\pm 0.023$ \\
11.0--12.5& $ 0.265\pm 0.082\pm 0.022$ \\
12.5--14.0& $ 0.187\pm 0.110\pm 0.019$ \\
14.0--24.0& $ 0.015\pm 0.005\pm 0.001$ \\
\hline
\end{tabular}
\end{table}

\begin{table}[]
\centering
\caption{Cross-section ratio between \jpsi-\psitwos and di-\jpsi production $\sigma_{\jpsi\text{-}\psitwos}/\sigma_{\text{di-}\jpsi}$ in intervals of $\Delta y$. The first uncertainty is statistical and the second systematic.}
\label{table:csRatioDY}
\begin{tabular}{cc}
\hline $\Delta y$ & $\sigma_{\jpsi\text{-}\psitwos}/\sigma_{\text{di-}\jpsi}$ \\\hline
0.0--2.2& $ 0.33\pm 0.12\pm 0.02$ \\
0.2--0.6& $ 0.17\pm 0.06\pm 0.01$ \\
0.6--1.0& $ 0.27\pm 0.07\pm 0.01$ \\
1.0--1.4& $ 0.49\pm 0.13\pm 0.03$ \\
1.4--1.8& $ 0.17\pm 0.12\pm 0.01$ \\
1.8--2.5& $ 0.36\pm 0.32\pm 0.05$ \\
\hline
\end{tabular}
\end{table}

\begin{table}[]
\centering
\caption{Cross-section ratio between \jpsi-\psitwos and di-\jpsi production $\sigma_{\jpsi\text{-}\psitwos}/\sigma_{\text{di-}\jpsi}$ in intervals of $\Delta\phi$. The first uncertainty is statistical and the second systematic.}
\label{table:csRatioDPHI}
\begin{tabular}{cc}
\hline $\Delta\phi/\pi$ & $\sigma_{\jpsi\text{-}\psitwos}/\sigma_{\text{di-}\jpsi}$ \\\hline
0--2& $ 0.36\pm 0.13\pm 0.02$ \\
2--4& $ 0.43\pm 0.10\pm 0.02$ \\
4--6& $ 0.08\pm 0.07\pm 0.00$ \\
6--8& $ 0.24\pm 0.10\pm 0.01$ \\
8--12& $ 0.24\pm 0.06\pm 0.01$ \\
\hline
\end{tabular}
\end{table}

\begin{table}[]
\centering
\caption{Cross-section ratio between \jpsi-\psitwos and di-\jpsi $\sigma_{\jpsi\text{-}\psitwos}/\sigma_{\text{di-}\jpsi}$ in intervals of $\pt^{\text{di-}\psi}$ [\ensuremath{\text{Ge\kern -0.1em V}/c}]. The first uncertainty is statistical and the second systematic.}
\label{table:csRatioPT}
\begin{tabular}{cc}
\hline $\pt^{\text{di-}\psi}$ [\ensuremath{\text{Ge\kern -0.1em V}/c}] & $\sigma_{\jpsi\text{-}\psitwos}/\sigma_{\text{di-}\jpsi}$ \\\hline
0--2& $ 0.23\pm 0.08\pm 0.01$ \\
2--4& $ 0.23\pm 0.06\pm 0.01$ \\
4--6& $ 0.39\pm 0.14\pm 0.02$ \\
6--8& $ 0.20\pm 0.08\pm 0.01$ \\
8--12& $ 0.35\pm 0.09\pm 0.02$ \\
12--24& $ 0.36\pm 0.10\pm 0.03$ \\
\hline
\end{tabular}
\end{table}

\begin{table}[]
\centering
\caption{Cross-section ratio between \jpsi-\psitwos and di-\jpsi $\sigma_{\jpsi\text{-}\psitwos}/\sigma_{\text{di-}\jpsi}$ in intervals of $y_{\text{di-}\psi}$. The first uncertainty is statistical and the second systematic.}
\label{table:csRatioY}
\begin{tabular}{cc}
\hline $y_{\text{di-}\psi}$ & $\sigma_{\jpsi\text{-}\psitwos}/\sigma_{\text{di-}\jpsi}$ \\\hline
2.0--2.5& $ 0.34\pm 0.18\pm 0.03$ \\
2.5--3.0& $ 0.31\pm 0.07\pm 0.01$ \\
3.0--3.5& $ 0.24\pm 0.06\pm 0.01$ \\
3.5--4.0& $ 0.24\pm 0.06\pm 0.01$ \\
4.0--4.5& $ 0.11\pm 0.09\pm 0.01$ \\
\hline
\end{tabular}
\end{table}

\clearpage


\addcontentsline{toc}{section}{References}
\bibliographystyle{LHCb}
\bibliography{main,standard,LHCb-PAPER,LHCb-CONF,LHCb-DP,LHCb-TDR}

\newpage
\centerline
{\large\bf LHCb collaboration}
\begin
{flushleft}
\small
R.~Aaij$^{35}$\lhcborcid{0000-0003-0533-1952},
A.S.W.~Abdelmotteleb$^{54}$\lhcborcid{0000-0001-7905-0542},
C.~Abellan~Beteta$^{48}$,
F.~Abudin{\'e}n$^{54}$\lhcborcid{0000-0002-6737-3528},
T.~Ackernley$^{58}$\lhcborcid{0000-0002-5951-3498},
B.~Adeva$^{44}$\lhcborcid{0000-0001-9756-3712},
M.~Adinolfi$^{52}$\lhcborcid{0000-0002-1326-1264},
P.~Adlarson$^{78}$\lhcborcid{0000-0001-6280-3851},
H.~Afsharnia$^{11}$,
C.~Agapopoulou$^{46}$\lhcborcid{0000-0002-2368-0147}
% ...

{\footnotesize \it

% ...
$^{11}$Universit{\'e} Clermont Auvergne, CNRS/IN2P3, LPC, Clermont-Ferrand, France\\
% ...
$^{35}$Nikhef National Institute for Subatomic Physics, Amsterdam, Netherlands\\
% ...
$^{44}$Instituto Galego de F{\'\i}sica de Altas Enerx{\'\i}as (IGFAE), Universidade de Santiago de Compostela, Santiago de Compostela, Spain\\
% ...
$^{46}$European Organization for Nuclear Research (CERN), Geneva, Switzerland\\
% ...
$^{48}$Physik-Institut, Universit{\"a}t Z{\"u}rich, Z{\"u}rich, Switzerland\\
% ...
$^{52}$H.H. Wills Physics Laboratory, University of Bristol, Bristol, United Kingdom\\
% ...
$^{54}$Department of Physics, University of Warwick, Coventry, United Kingdom\\
% ...
$^{57}$School of Physics and Astronomy, University of Glasgow, Glasgow, United Kingdom\\
$^{58}$Oliver Lodge Laboratory, University of Liverpool, Liverpool, United Kingdom\\
% ...
$^{78}$Department of Physics and Astronomy, Uppsala University, Uppsala, Sweden, associated to $^{57}$\\
% ...
}
\end{flushleft}

\end{document}